\documentclass[11pt,a4paper]{article}
\usepackage{amsmath,amssymb,amsthm,eucal,cite,color}
\usepackage[top=3cm,right=3cm,left=2.5cm,bottom=3cm]{geometry}
\usepackage{multirow}

\title{The Master Ward Identity for scalar QED}

\author{Michael D\"utsch, Luis Peters, Karl-Henning Rehren
\thanks{Email: 
\texttt{michael.duetsch@theorie.physik.uni-goettingen.de,
luis.peters@stud.uni-goettingen.de, krehren@gwdg.de}} \\[2mm] Institute for
Theoretical Physics \\ Georg-August University G\"ottingen \\ Friedrich-Hund-Platz 1,
37077 G\"ottingen, Germany}

\date{\today}

\newcommand{\be}{\begin{equation}}   
\newcommand{\ee}{\end{equation}} 

\newcommand{\al}{\alpha}          
\newcommand{\bt}{\beta}           
\newcommand{\Dl}{\Delta}          
\newcommand{\dl}{\delta}          
\newcommand{\eps}{\varepsilon}    
\newcommand{\ka}{\kappa}          
\newcommand{\La}{\Lambda}         
\newcommand{\la}{\lambda}         
\newcommand{\om}{\omega}          
\renewcommand{\th}{\theta}        
\newcommand{\vf}{\varphi}         

\newcommand{\bC}{\mathbb{C}}      
\newcommand{\bM}{\mathbb{M}}      
\newcommand{\bN}{\mathbb{N}}      
\newcommand{\bR}{\mathbb{R}}      
\newcommand{\bZ}{\mathbb{Z}}      

\newcommand{\nS}{\mathbf{S}}      

\newcommand{\sA}{\mathcal{A}}     
\newcommand{\sC}{\mathcal{C}}     
\newcommand{\sD}{\mathcal{D}}     
\newcommand{\sF}{\mathcal{F}}     
\newcommand{\sO}{\mathcal{O}}     
\newcommand{\sP}{\mathcal{P}}     
\newcommand{\sR}{\mathcal{R}}     
\newcommand{\sT}{\mathcal{T}}     
\newcommand{\sY}{\mathcal{Y}}
\newcommand{\sZ}{\mathcal{Z}}

\newcommand{\homog}{\mathrm{hom}} 
\newcommand{\loc}{\mathrm{loc}}   

\newcommand{\bigveestar}{\bigvee\nolimits_{\!\star}} 
\newcommand{\del}{\partial}       

\newcommand{\longto}{\longrightarrow}
\newcommand{\less}{\setminus}     
\newcommand{\ovl}{\overline}      
\newcommand{\ox}{\otimes}         
\newcommand{\oxsym}{\otimes_\mathrm{s}}         
\newcommand{\bigox}{\bigotimes}         
\newcommand{\oxyox}{\otimes\cdots\otimes} 
\newcommand{\unl}{\underline}     
\newcommand{\up}{{\mathord{\uparrow}}} 
\newcommand{\wh}{\widehat}        
\newcommand{\x}{\times}           
\renewcommand{\:}{\colon}         

\DeclareMathOperator{\sd}{sd}      
\DeclareMathOperator{\supp}{supp}  

\newcommand{\ldbrack}{[\mskip-2.5mu[} 
\newcommand{\rdbrack}{]\mskip-2.5mu]} 

\newcommand{\pw}[1]{\ldbrack#1\rdbrack} 
\newcommand{\set}[1]{\{\,#1\,\}}    
\newcommand{\word}[1]{\quad\mbox{#1}\quad} 
\def\wick:#1:{\mathopen:#1\mathclose:} 

\newcommand{\fd}[2]{\frac{\dl#1}{\dl\phi(#2)}} 
\def\duo<#1,#2>{\langle#1,#2\rangle} 

\numberwithin{equation}{section}

\theoremstyle{plain}
\newtheorem{thm}{Theorem}[section]  
\newtheorem{defn}[thm]{Definition}  

\theoremstyle{definition}
\newtheorem{remk}[thm]{Remark}      

\newtheoremstyle{example}
   {\topsep}{\topsep}{\small}{0pt}%
   {\bfseries}{.}{ }{}
\theoremstyle{example}

\newtheoremstyle{exercise}
   {\topsep}{\topsep}{\small}{0pt}%
   {\bfseries}{.}{ }{}
\theoremstyle{exercise}

\theoremstyle{remark}

\DeclareRobustCommand{\qned}{\ifmmode
  \else \leavevmode\unskip\penalty9999 \hbox{}\nobreak\hfill \fi
  \quad\hbox{\qnedsymbol}}
\newcommand{\qnedsymbol}{$\boxminus$} 

\makeatletter
\renewcommand{\section}{\@startsection{section}{1}{\z@}%
                        {-3.5ex \@plus -1ex \@minus -.2ex}%
                        {2.3ex \@plus.2ex}%
                        {\normalfont\large\bfseries}}
\renewcommand{\subsection}{\@startsection{subsection}{2}{\z@}%
                        {-3.25ex \@plus -1ex \@minus -.2ex}%
                        {1.5ex \@plus .2ex}%
                        {\normalfont\normalsize\bfseries}}
\renewcommand{\subsubsection}{\@startsection{subsubsection}{3}{\z@}%
                        {-3.25ex \@plus -1ex \@minus -.2ex}%
                        {1.5ex \@plus .2ex}%
                        {\normalfont\normalsize\bfseries}}
\renewcommand{\@dotsep}{200} 
\makeatother

\def\tilde{\widetilde}

\begin{document}

\maketitle

\begin{abstract}
It is emphasized that for interactions with derivative couplings, 
the Ward Identity (WI) securing the preservation of a global (1) 
  symmetry should be modified. Scalar QED is taken as an explicit
  example. More precisely, it is rigorously
shown in scalar QED that the naive WI and
  the improved Ward Identity (``Master Ward Identity'', MWI) are
  related to each other by a finite renormalization of the time-ordered 
product (``$T$-product'') for the derivative fields; and we point 
out that the MWI has advantages over the naive WI -- in 
particular with regard to the proof of the MWI. We 
show that the MWI can be fullfilled
in all orders of perturbation theory by an 
appropriate renormalization of the $T$-product, without conflict with 
other standard renormalization conditions.
Relations with other recent formulations of the MWI are established.

  \end{abstract}

\section{Introduction}

In spinor QED the Master Ward Identity (MWI) expressing global 
$U(1)$-symmetry contains all information that is needed for 
a consistent perturbative BRST-construction
of the model, see \cite{DF99} or \cite[Chap.~5]{D19}. This ``QED-MWI'' 
is a renormalization condition on $T$-products\footnote{We understand the term
`renormalization condition' in the precise sense of the inductive Epstein-Glaser construction of
$T$-products \cite{EG73}: a constraint on extensions of
distributions, as explained in Appendix \ref{sec:axioms-T}.} 
 to be satisfied to 
all orders of perturbation theory. It reads
\be\label{eq:MWI-QED}
\del_y^\mu\,T_{n+1}\bigl(\tilde B_1(x_1)\ox\cdots\ox \tilde B_n(x_n)\ox
j_\mu(y)\bigr)_0=
-\sum_{l=1}^n\dl(y-x_l)\,T_n\bigl(\tilde B_1(x_1)\ox \cdots
\ox\widetilde{\th B_l}(x_l)\ox\cdots\ox \tilde B_n(x_n)\bigr)_0,
\ee
where $j^\mu = \ovl{\psi}\gamma^\mu\psi$ is the
Dirac current, $B_1,\ldots,B_n$ are arbitrary submonomials (see \eqref{eq:submonomials}) of the interaction
$L=e\,j^\mu A_\mu$, and $\th$ is the charge number operator. The notation 
$\widetilde{(\cdot)}$ means that fermionic field polynomials
are converted into bosonic field polynomials by multiplying them with a
Grassmann variable.
By $T(\dots)_0$ we denote on-shell $T$-products (see below).

There is an essential difference between spinor QED and scalar QED:
in the latter, the current to which the electromagnetic potential is 
coupled, contains \emph{first derivatives} of the basic fields:%
\footnote{This is the Noether current pertaining to the invariance 
of the free action of the scalar field under the global $U(1)$-transformation
$\phi(x)\to e^{i\al}\phi(x)$ ($\al\in\bR$). 
The Dirac current is defined w.r.t.\
$\psi \to e^{-i\alpha}\psi$. This switch of sign convention will explain
a number of opposite signs in the present formulas as compared
to spinor QED in \cite{D19}, notably (\ref{eq:WI-hat}) and (\ref{eq:Q}).}
\be\label{eq:sc-curr}
j^\mu:=i(\phi\del^\mu\phi^*-\phi^*\del^\mu\phi).
\ee

\medskip

It is apriori not evident how to translate the QED-MWI \eqref{eq:MWI-QED} to
models with derivative couplings, and scalar QED may serve as a prototype of 
such models.

\medskip

Our results can be summarized as follows: in Sect.~\ref{sec:simple-WI} we
postulate a naive WI for scalar QED, just by analogy to spinor QED.
To fulfil it, an ``unnatural'' renormalization of the
$T$-product of $\del^\mu\phi(x)$ with
$\del^\nu\phi^*(y)$ is required 
\cite{DKS93}: one has to add $ig^{\mu\nu}\dl(x-y)$ to
$\del^\nu\del^\mu\Dl^F(x-y)$. This addition violates the standard
renormalization conditions `Field Equation' and `Action Ward Identity'.

In Sect.~\ref{sec:MWI} we work out the MWI 
for the global $U(1)$-transformation $\phi(x)\to e^{i\al}\phi(x)$ 
in scalar QED, and find that, compared with the naive WI, it 
contains an additional term.

In Sect.~\ref{sec:MWI-proof} we prove that the MWI can be fulfilled by an 
appropriate renormalization of the $T$-product, which is compatible with the
further standard renormalization conditions.

In Sect.~\ref{sec:MWI->WI}, starting with the time-ordered product ``$T$'', 
we define in all orders a new time-ordered
product $\wh T$ induced from the initial finite renormalization 
$\del^\nu\del^\mu\Dl^F\to\del^\nu\del^\mu\Dl^F+ig^{\mu\nu}\dl$, 
by the inductive Epstein-Glaser method \cite{EG73}. We
prove that the validity of the MWI for $T$ is equivalent to
the validity of naive WI for $\wh T$. 
In fact, one may continuously interpolate between $T$ and $\wh T$.

In Sect.~\ref{sec:unitary-MWI} we prove, in the perturbative approach 
to scalar QED, that the MWI is equivalent to the so-called ``unitary MWI''. 
The latter is an identity, conjectured by Fredenhagen \cite{BDFR20}, which 
seems to be well suited for the formulation of symmetries in the 
Buchholz-Fredenhagen quantum algebra \cite{BF19}.

All proofs are given to all orders of perturbation theory.

\subsection{Some technical preparations}

We use natural units, in particular $\hbar=1$, and the underlying spacetime
is the $4$-dimensional Minkowski space $\bM$. We work with causal
perturbation theory,  also called `Epstein-Glaser method'
\cite{EG73}. This method is based on an axiomatic definition of the time-ordered product $T\equiv (T_n)_{n=1}^\infty$, 
the most important axiom being a causal factorization property of $T_n$
(see Appendix \ref{sec:axioms-T}), and yields an inductive construction of the
sequence $(T_n)$ solving the axioms. In addition, we use
the formalism where quantum fields are functionals on classical configuration
  spaces, equipped with a non-commutative product: the star product of 
    the free theory (denoted by ``$\star$'', see \eqref{eq:star-product}). Perturbation theory
      represents interacting fields as formal power series within this algebra,
 using the time-ordered product of local fields, which is commutative. The prominent mathematical
task is the construction of the time-ordered product.
For details and conventions, we refer to
  the book \cite{D19}, where in particular the conventions for the
  propagators are fixed in \cite[App.~A.2]{D19}. 

For the convenience of the reader, we sketch some basic definitions of the 
formalism for the model at hand, that is, scalar QED. The expert reader may skip the remainder 
of this section, except for the third and second last paragraph containing some remarks about
`on-shell MWI versus off-shell MWI' and the definition of $\sP$.
The basic fields of scalar QED are a
complex scalar field $\phi(x)$, its conjugate field $\phi^*(x)$ and the photon field 
$A(x)\equiv(A^\mu(x))$.
The configuration space is $\sC=C^\infty(\bM,\bC)\x C^\infty(\bM,\bR^4)$,
where the first factor stands for the configurations of $\phi,\phi^*$ and the 
second for the configurations of $(A^\mu)$. The basic fields are the evaluation functionals
$$
\phi(x)[h]=h(x),\quad \phi^*(x)[h]=\ovl{h(x)},\quad A^\mu(x)[a]=a^\mu(x),\quad
\forall h\in C^\infty(\bM,\bC),\,\,a\equiv (a_\mu)\in C^\infty(\bM,\bR^4),
$$
where the overline denotes complex conjugation. 
The \emph{space of fields} $\sF$ is the set of all polynomial 
functionals on the configuration space satisfying certain properties. More 
precisely, a field $\sF\ni F\: \sC \to \bC$ is a \emph{finite} sum of functionals of the form
\begin{align}
F =& \sum_{p,n,l} \int dx_1 \cdots dx_p\,dy_1\cdots dy_n\,dz_1\cdots dz_l \,\,
\prod_{i=1}^p A_{\mu_i}(x_i)\,\prod_{j=1}^n\phi(y_j)\,\prod_{k=1}^l\phi^*(z_k)
\nonumber\\
&\cdot f_{p,n,l}^{\mu_1\ldots \mu_p}(x_1,\dots,x_p,y_1,\ldots,y_n,z_1,\ldots,z_l)
\nonumber\\
=:& \sum_{p,n,l}\big\langle f_{p,n,l}^{\mu_1\ldots \mu_p},(\ox_{i=1}^p 
A_{\mu_i})\ox\phi^{\ox n}\ox(\phi^*)^{\ox l}\big\rangle,\label{eq:F}
\end{align}
evaluated as
$$
F[h,a]:= \sum_{p,n,l}\big\langle f_{p,n,l}^{\mu_1\ldots \mu_p},
(\ox_{i=1}^p a_{\mu_i})\ox h^{\ox n}
\ox(\ovl h)^{\ox l}\big\rangle\quad\forall (h,a)\in \sC,
$$
where $f_{0,0,0}\in\bC$ is constant; and for $p+n+l \geq 1$, each expression 
$f_{p,n,l}^{\mu_1\ldots}$ is an element of 
$\sD'(\bM^{p+n+l},\bC)$ with compact support, 
which satisfies a certain wave front set condition (not relevant in this work, see 
\cite[Def.~1.2.1]{D19}). The purpose of the latter is to ensure the existence of the 
pointwise products of distributions appearing in the definition of the star product \eqref{eq:star-product}.

The support of $F\in\sF$ is defined by
\be\label{eq:supp}
\supp F:=\ovl{\bigcup_{(h,a)\in\sC}\,\,\bigcup_{\vf=A^\mu,\phi,\phi^*}\supp\frac{\dl F}{\dl\vf(\cdot)}[h,a]},
\ee
where on the r.h.s.~we mean the support in the sense of distributions.

Convergence in $\sF$ is understood in the pointwise sense: $\lim_{n\to\infty}F_n=F$ if and only if
$\lim_{n\to\infty}F_n[h,a]=F[h,a]$ for all $(h,a)\in \sC$ (cf.~\cite[formula (1.2.3)]{D19}). 
For example, the closure on the r.h.s.~of \eqref{eq:supp} is done in this sense; or, by $\sD'(\bM,\sF)$ we mean 
the vector space of all linear maps from $\sD(\bM)$ to $\sF$, which are continuous w.r.t.~this topology on $\sF$.
 
The vacuum expectation value (VEV) of a field $F\in\sF$ is $\om_0(F):=F[0,0]$.

For the complex scalar field, the free field equation is the Klein-Gordon equation: 
$(\square+m^2)\phi(x)=0=(\square+m^2)\phi(x)^*$. For the photon field
we use the Feynman gauge, 
that is, the free field equation is the wave equation $\square A^\mu(x)=0$; see \cite[Sect.~5.1.3]{D19}.

The space of fields $\sF$ is equipped with the following operations:
\begin{itemize}
\item An involutive \emph{$*$-operation}, defined on the elements \eqref{eq:F} by
$$
F^*:= \sum_{p,n,l}\big\langle \ovl{f_{p,n,l}^{\mu_1\ldots \mu_p}},(\ox_{i=1}^p 
A_{\mu_i})\ox(\phi^*)^{\ox n}\ox\phi^{\ox l}\big\rangle
$$
(again the overline denotes complex conjugation), in particular $\phi$ and $\phi^*$ are mutually exchanged;
\item the \emph{pointwise} or \emph{classical product},
$$
(F\cdot G)[h,a]:=F[h,a]\cdot G[h,a],\quad\forall F,G\in\sF,\,\,(h,a)\in C^\infty(\bM,\bC)\x C^\infty(\bM,\bR^4),
$$
which is \emph{commutative}.
\item The free theory is quantized by deforming the classical product into a noncommutative product -- the \emph{star product};
to simplify the notations we give here the definition for the model of
one real scalar field $\vf$:
\begin{align}\label{eq:star-product}
 F \star G :=
\sum_{n=0}^\infty \frac{\hbar^n}{n!} 
\int dx_1 \cdots dx_n\,dy_1 \cdots dy_n\, 
\frac{\dl^n F}{\dl\vf(x_1)\cdots\dl\vf(x_n)}
\prod_{l=1}^n \Dl^+_m(x_l - y_l) \,
\frac{\dl^n G}{\dl\vf(y_1)\cdots\dl\vf(y_n)} \,,
\end{align}
where $\Delta^+_m$ is the Wightman two-point function to the mass $m$. Exceptionally, we write here $\hbar$, because it is
the deformation parameter. More precisely, the star product is a map $\sF\pw{\hbar} \x \sF\pw{\hbar} \to \sF\pw{\hbar}$, where 
$\sF\pw{\hbar}$ denotes the space of formal power series in $\hbar$ with coefficients in $\sF$.
When \eqref{eq:star-product} is adapted to scalar QED, the only non-vanishing `contractions' in the definition of the star product are
$$
\om_0\bigl(\phi^*(x)\star\phi(y)\bigr)=\hbar\,\Dl^+_m(x-y)= \om_0\bigl(\phi(x)\star\phi^*(y)\bigr),\quad
\om_0\bigl(A^\mu(x)\star A^\nu(x)\bigr)=-\hbar g^{\mu\nu}\,D^+(x-y),
$$
where $D^+:=\Dl^+_{m=0}$.
\end{itemize}

To a far extent, we work \emph{on-shell}.
This means that all functionals $F\in\sF$ are restricted
to the space $\sC_{S_0}$ of solutions of the free field equations; 
we indicate this restriction by
$$
F_0:=F\big\vert_{\sC_{S_0}}\quad\forall F\in\sF.
$$
Algebraically, on-shell fields can be identified with Fock space operators,
where the star product of on-shell fields 
corresponds to the operator product, and the pointwise 
product of on-shell functionals (i.e.,
$(F_0\cdot G_0)[h,a]:=F_0[h,a]\cdot G_0[h,a]$ 
for all $F,G\in\sF$ and $(h,a)\in\sC_{S_0}$)
to the normally ordered product, see \cite[Thm.~2.6.3]{D19}. The motivations 
to study only the on-shell version of the MWI in this paper are the following: firstly, the essential 
information of the off-shell MWI is already contained in its on-shell version (see Remark \ref{rm:off-shellMWI}); 
secondly, working in Fock space one ``sees'' only the on-shell MWI;
and finally, off-shell notation would just unnecessarily overburden many formulas.  

Throughout this paper, we need only distributions $f_{p,n,l}^{\mu_1\ldots}$
such that no derivatives of $A^\mu$ and solely zeroth and first derivatives of
$\phi$ and $\phi^*$ appear. So we define $\sP$ to be the space of polynomials 
in $A^\mu,\phi,\phi^*,\del^\mu\phi$ and $\del^\nu\phi^*$ only.

The subspace $\sF_\loc\subset\sF$ of \emph{local} fields is the linear 
span of the set
$\set{B(g)\equiv\int dx \,\,g(x)\,B(x)\,\big\vert\,B\in\sP,\,g\in\sD(\bM)}$.
For example: 
$$
(\del^\mu\phi^*\del_\mu\phi)(g)[h,a]=\int dx\,\,g(x)\,
\del^\mu\ovl{h(x)}\del_\mu h(x),\quad\forall (h,a)\in\sC.
$$

\section{The naive Ward Identity}
\label{sec:simple-WI}

A natural candidate for the Ward Identity (WI) expressing global 
$U(1)$-symmetry for scalar QED just copies the QED-MWI
\eqref{eq:MWI-QED} with the charge number operator
\be\label{eq:theta}
\th B:=\phi\,\frac{\del B}{\del\phi}+
\del^\mu\phi\,\frac{\del B}{\del(\del^\mu\phi)}
-\phi^*\,\frac{\del B}{\del\phi^*}-
\del^\mu\phi^*\,\frac{\del B}{\del(\del^\mu\phi^*)}
\word{for $B\in\sP$,}
\ee
and with the time-ordered product $\wh T$
that is required to satisfy the basic axioms (i)-(iv) and the 
renormalization conditions (v)-(viii) listed in Appendix
\ref{sec:axioms-T}.
This yields 
\begin{align}\label{eq:WI-hat}
\del_y^\mu\,\wh T_{n+1}\bigl(&B_1(x_1)\ox\cdots\ox  B_n(x_n)\ox
j_\mu(y)\bigr)_0=\nonumber\\
&\sum_{l=1}^n\dl(y-x_l)\,\wh T_n\bigl( B_1(x_1)\ox \cdots
\ox(\th B_l)(x_l)\ox\cdots\ox  B_n(x_n)\bigr)_0,
\end{align}
where $B_1,\ldots,B_n$ are arbitrary submonomials of the interaction
\be\label{eq:tilde-L}
\tilde L:=e\,j^\mu A_\mu .
\ee
This Ward identity is a generalization of the one postulated and proved in
\cite{DKS93}; the difference is that in this reference only
  the neutral fields $B_1,\dots,B_n\in\{\tilde L,j^\mu\}$ are studied and, hence, the r.h.s.~of
\eqref{eq:WI-hat} vanishes. In \cite{DKS93},
the WI there is motivated by gauge invariance of the 
on-shell $S$-matrix (i.e., the restriction to $\sC_{S_0}$ of the $S$-matrix defined 
in \eqref{eq:S-matrix}), that is, invariance under the 
transformation $A^\mu(x)\to A^\mu(x)+\del^\mu\La(x)$ of $\wh\nS(g,ej^\mu A_\mu)_0$ 
in the formal adiabatic limit $g(x)\to 1\,\,\forall x$. 
The task would be to establish the existence of $\wh T$
satisfying (i)-(viii) and \eqref{eq:WI-hat}.

Below in Sect.~\ref{sec:MWI} we show that \eqref{eq:WI-hat} is
only a simplified 
version of the Master Ward Identity (MWI) expressing $U(1)$-symmetry; 
the latter is better suited for models with derivative couplings,
and is easier to establish.

Particular cases of the WI \eqref{eq:WI-hat} are
\begin{align}\label{eq:WI-hat-2}
&\del^y_\mu\,\wh T\bigl(j^\nu(x)\ox j^\mu(y)\bigr)_0=0,\quad
\del^y_\mu\,\wh T\bigl(\del^\nu\phi(x)\ox j^\mu(y)\bigr)_0=
\,\dl(y-x)\,\del^\nu\phi(x)_0,\nonumber\\
&\del^y_\mu\,\wh T\bigl(\del^\nu\phi^*(x)\ox j^\mu(y)\bigr)_0=
-\,\dl(y-x)\,\del^\nu\phi^*(x)_0.
\end{align}
These identities have an important property: 
Requiring that $\wh T$ satisfies the axiom (v) Field Independence, that is,
the validity of the causal Wick expansion \eqref{eq:causal-Wick-expan}, 
the tree diagram part of the first identity, and 
the other two identities are fulfilled if and only if the numerical
  distribution  $\hat t(\del^\nu\phi,\del^\mu\phi^*)=\omega_0(\wh
      T(\del^\nu\phi\otimes \del^\mu\phi^*))$ (cf.\ \eqref{eq:tn})
is specified as
\be\label{eq:ddDF+dl}
\hat t(\del^\nu\phi,\del^\mu\phi^*)(x-y)=
-\del^\nu\del^\mu\Dl^F(x-y)-ig^{\mu\nu}\dl(x-y)
=\hat t(\del^\nu\phi^*,\del^\mu\phi)(x-y),
\ee
as one sees by explicit computation. The finite renormalization
of the Feynman propagator with two derivatives
\be\label{eq:fin-ren}
\del^\nu\del^\mu\Dl^F(x-y)\longmapsto
\del^\nu\del^\mu\Dl^F(x-y)+ig^{\mu\nu}\dl(x-y)
\ee
is admissible in the framework of causal perturbation theory,
since the singular order is $\om(\del^\nu\del^\mu\Dl^F)=0$ 
(see \eqref{eq:nonunique} for the definition of
the singular order).

The additional term $ig^{\mu\nu}\dl(x-y)$ has the advantage, that it generates 
as a necessary finite ``counter term'' the quartic interaction part, i.e.,
$e^2\,A^\mu A_\mu\,\phi\phi^*$ (as it was first realized in
\cite{DKS93}), propagating correctly to higher orders in the inductive 
Epstein--Glaser construction of $\wh T\equiv (\wh T_n)$. Indeed, for the 
$S$-matrix belonging to $\wh T$ (defined in \eqref{eq:S-matrix}) we obtain
\begin{align}
\wh\nS(&g,ej A)= 1+ie(j A)(g)\nonumber\\
&-\frac{e^2}2\int dx\,dy\,\,g(x)g(y)\bigl[
\hat t(\del^\mu\phi^*,\del^\nu\phi)(x-y)\,A_\mu(x)\phi(x)\,A_\nu(y)\phi^*(y)
+(\phi\leftrightarrow\phi^*)\bigr]+\ldots\nonumber\\
=&1+i\bigl(e\,(jA)(g)+e^2\,(AA\phi^*\phi)(g^2)\bigr)+\ldots,\label{eq:L-ren}
\end{align}
where the dots contain further terms of order $\sO((eg)^2)$ and all terms of
higher orders in $(eg)$.
 
But the addition $ig^{\mu\nu}\dl(x-y)$
has the disadvantages that it violates the renormalization condition `Field Equation' (FE) and 
the `Action Ward Identity' (AWI) (generally formulated in Appendix 
\ref{sec:axioms-T}):
\begin{align*}
&\word{FE:}
\hat t(\del^\nu\phi,\del^\mu\phi^*)(x-y)\not=\int dz\,\,\del^\nu\Dl^F(x-z)\,
\frac{\dl\,\del^\mu\phi^*(y)}{\dl\phi^*(z)}\Bigl(=-\del^\nu\del^\mu\Dl^F(x-y)
\Bigr),\\
&\word{AWI:}
\hat t(\del^\nu\phi,\del^\mu\phi^*)(x-y)\not=
\del^\nu_x\del^\mu_y\hat t(\phi,\phi^*)(x-y)\Bigl(=-\del^\nu\del^\mu\Dl^F(x-y)
\Bigr).
\end{align*}
A proof of the WI \eqref{eq:WI-hat} along the lines of the proof of the
QED-MWI in \cite[Chap.~5.2.2]{D19} would require additional work, because
that proof uses essentially that the time-ordered product fulfills the
``Field Equation''. Instead, an indirect proof via the 
  MWI will be given in Sect.\ \ref{sec:MWI->WI}.

\section{The Master Ward Identity}
\label{sec:MWI}

Due to the mentioned bad properties of  the time-ordered product $\wh T$ 
and the resulting problems in trying to adapt the proof of the QED-MWI to the 
WI \eqref{eq:WI-hat}, we prefer to work with the complete
relevant MWI for scalar QED.

The original references for the MWI are \cite{DB02,DF03} and \cite{BD08}.
It is a universal formulation of symmetries; it can be understood as 
the straightforward generalization to QFT of the most general classical 
identity for local fields that can be obtained from the field equation 
and the fact that classical fields may be multiplied
pointwise. In contrast, the quantum version of the MWI is a renormalization
condition with regard to the axioms for the $T$-product (cf.~Appendix 
\ref{sec:axioms-T}). It cannot 
always be fulfilled due to the well-known anomalies.

\subsection{Working out the relevant MWI for scalar QED}

Generally, the on-shell MWI (see \cite{DF03,BD08} and 
\cite[Chap.~4.2]{D19}) is derived from the symmetry at hand. It reads
\begin{align}\label{eq:MWI-general}
T_{n+1}\bigl(B_1(x_1)&\ox\cdots\ox  B_n(x_n)\ox \dl_{Q(y)}S_0\bigr)_0=
\nonumber\\
& i\sum_{l=1}^n T_n\bigl( B_1(x_1)\ox \cdots
\ox \dl_{Q(y)}B_l(x_l)\ox\cdots\ox  B_n(x_n)\bigr)_0,
\end{align}
where $\dl_{Q(y)}$ is a functional differential operator specified by the 
symmetry, and the time-ordered product $T$ is required to fulfil the axioms
(i)-(viii) given in Appendix \ref{sec:axioms-T} and the additional 
renormalization conditions AWI and FE.

In the case at hand, we study the global $U(1)$-transformation 
$\phi(y)\to e^{i\al}\phi(y)$ ($\al\in\bR$). Let
\be\label{eq:Q}
Q(y):=-\frac{d}{d\al}\Big\vert_{\al=0}e^{i\al}\phi(y)=-i\phi(y)
\ee
and the pertinent functional differential operator
\be\label{eq:delta-Q}
\dl_{Q(y)}:=Q(y)\,\frac{\dl}{\dl\phi(y)}+Q^*(y)\,\frac{\dl}{\dl\phi^*(y)}.
\ee
Introduce a modification $\th_\mu$ of the charge number operator,
\be\label{eq:th-mu}
\th_\mu B:=\phi\,\frac{\del B}{\del(\del^\mu\phi)}
-\phi^*\,\frac{\del B}{\del(\del^\mu\phi^*)}
\word{for $B\in\sP$,}
\ee
and recall that%
\footnote{$S_0$ is a formal expression
      that is not an element of $\sF$, because the configurations are
    not necessarily vanishing at infinity. Only functional derivatives of $S_0$ are really needed, e.g., 
$\frac{\dl S_0}{\dl\phi^*(x)}=-(\square+m^2)\phi(x)$, and the latter are well defined functionals.} 
\be\label{eq:S0}
S_0:=\int dx \bigl(\del_\mu\phi^*(x)\del^\mu\phi(x)-m^2\phi^*(x)\phi(x)\bigr)+S_0(A^\mu).
 \ee
Then, one verifies straightforwardly that
\be\label{eq:dl_Q}
\dl_{Q(y)}\,S_0=\del_\mu j^\mu(y),\quad
\dl_{Q(y)}B(x)=-i\Bigl(\dl(y-x)\,(\th B)(x)-
\del_y^\mu\bigl(\dl(y-x)\,(\th_\mu B)(x)\bigr)\Bigr).
\ee
For scalar QED and the symmetry given by the above defined $Q$,
the MWI takes the particular form (cf.~\cite[Exer.~4.2.6]{D19})
\begin{align}\label{eq:MWI}
\del_y^\mu\,T_{n+1}\bigl(B_1(x_1)&\ox\cdots\ox  B_n(x_n)\ox
j_\mu(y)\bigr)_0=\nonumber\\
&\sum_{l=1}^n\dl(y-x_l)\,T_n\bigl( B_1(x_1)\ox \cdots
\ox(\th B_l)(x_l)\ox\cdots\ox  B_n(x_n)\bigr)_0
\nonumber\\
&-\del^\mu_y\Bigl(\sum_{l=1}^n\dl(y-x_l)\,T_n\bigl( B_1(x_1)\ox \cdots
\ox(\th_\mu B_l)(x_l)\ox\cdots\ox  B_n(x_n)\bigr)_0\Bigr)
\end{align}
for $B_1,\ldots,B_n\in\sP$, by using the AWI.
Compared with \eqref{eq:WI-hat}, the additional terms (i.e., the terms in the
last line) arise from the last term in the formula \eqref{eq:dl_Q}
for $\dl_{Q(y)}B(x)$.

Instead of the identities \eqref{eq:WI-hat-2} we now obtain
\begin{align}\label{eq:MWI-2}
&\del^y_\mu\,T_2\bigl(\del^\nu\phi(x)\ox j^\mu(y)\bigr)_0=
\dl(y-x)\,\del^\nu\phi(x)_0-(\del^\nu\dl)(y-x)\,\phi(x)_0,\nonumber\\
&\del^y_\mu\,T_2\bigl(\del^\nu\phi^*(x)\ox j^\mu(y)\bigr)_0=
-\dl(y-x)\,\del^\nu\phi^*(x)_0+(\del^\nu\dl)(y-x)\,\phi^*(x)_0,\nonumber\\
&\del^y_\mu\,T_2\bigl(j^\nu(x)\ox j^\mu(y)\bigr)_0=
2i\,(\phi^*\phi)(x)_0\,\del^\nu\delta(y-x),
\end{align}
by using $\th^\mu j^\nu=-2ig^{\mu\nu}\,\phi\phi^*$.

When working with the time-ordered product $T$ 
satisfying the MWI \eqref{eq:MWI}
one has to add the quartic interaction part ``by hand'', that is, one starts 
the inductive Epstein-Glaser construction of the $S$-matrix with 
the following interaction $S$:
\be\label{eq:interaction}
T_1(S)=S:=e\,(j^\mu A_\mu)(g)+e^2\,(A^\mu A_\mu\phi^*\phi)(g^2)\in\sF_\loc.
\ee
The addition of the quartic interaction term can be motivated by classical 
gauge invariance. In this procedure, the order of the time-ordered product does 
not agree with the order in the coupling constant $(eg)$; 
gauge invariance of the $S$-matrix must hold in each order in $(eg)$ 
individually.  

\begin{remk}\label{rm:off-shellMWI}[Off-shell MWI] The off-shell MWI differs from the on-shell MWI 
by additional terms proportional to the field equation for $\phi$ and $\phi^*$. In detail, the off-shell MWI for 
scalar QED is obtained from the on-shell MWI \eqref{eq:MWI} by omitting the restriction of all $T$-products to 
$\sC_{S_0}$ and by adding on the r.h.s.~the two terms
\begin{align*}
&+i\, T_{n+1}\bigl(B_1(x_1)\ox\cdots\ox  B_n(x_n)\ox \phi(y)\bigr)\cdot(\square+m^2)\phi^*(y)\\
&-i\, T_{n+1}\bigl(B_1(x_1)\ox\cdots\ox  B_n(x_n)\ox \phi^*(y)\bigr)\cdot(\square+m^2)\phi(y).
\end{align*}
The proof of the on-shell MWI given in Sect.~\ref{sec:MWI-proof} can be extended to the off-shell MWI by
very minor supplements, as one sees by comparing with the proof of the off-shell MWI for spinor QED given in
\cite[Sect.~5.2.2]{D19}.
\end{remk}

\section{Equivalent reformulations of the MWI}

Some remarks on the notations: in this section we solely work with the 
time-ordered product $T\equiv (T_n)$, which satisfies the AWI. Thus we may 
interpret $T_n$ as a map $T_n:\sF_\loc^{\oxsym n}\to\sF$, and for the $S$-matrix 
\eqref{eq:S-matrix} we may write 
\be\label{eq:S-matrix-1}
\nS(F)\equiv T\bigl(e_{\oxsym}^{iF}\bigr):=
1+\sum_{n=1}^\infty\frac{i^n}{n!}\,T_n(F^{\oxsym n})
\ee
in the sense of formal power series in $F$, 
where $\oxsym$ denotes the symmetrized tensor product. In addition,
let $B\in\sP$ and $g,\al\in\sD(\bM,\bR)$ the function switching the coupling 
constant and an infinitesimal local $U(1)$-transformation, respectively.

\subsection{The MWI as an identity for formal power series}

Motivated by the expressions 
\begin{align*}
&\int dy\,dx\,\,\al(y)\,g(x)\,\dl(y-x)\,(\th B)(x)=(\th B)(g\al),\\
&-\int dy\,dx\,\,\al(y)\,g(x)\,\del^\mu_y\dl(y-x)\,(\th_\mu B)(x)=
(\th_\mu B)(g\del^\mu\al),
\end{align*}
which appear in the MWI \eqref{eq:MWI} when integrated out with 
$\al(y)\,\prod_jg_j(x_j)\in\sD(\bM^{n+1},\bR)$, 
we introduce two derivations (i.e., linear maps 
satisfying the Leibniz rule) on $\sT(\sF_\loc)$ (by which we mean the
the linear space spanned by the sequences $(F_{k,1}\oxsym\cdots\oxsym F_{k,k})_{k=0}^\infty$ 
in the tensor algebra on $\sF_\loc$, formally written as $\sum_{k=0}^\infty (F_{k,1}\oxsym\cdots\oxsym F_{k,k})$):
\begin{align}\label{eq:dl,d}
&\dl^{(0)}_\th(\al),\,\dl^{(1)}_\th(\al):\,\sT(\sF_\loc)\longto\sT(\sF_\loc)
\word{uniquely specified by}\nonumber\\
&\dl^{(0)}_\th(\al)\bigl(B(g)\bigr):=(\th B)(g\al)\word{and}
\dl^{(1)}_\th(\al)\bigl(B(g)\bigr):=(\th_\mu B)(g\del^\mu\al),
\word{respectively.}
\end{align}
An immediate consequence is the relation
$$
d\bigl(e_{\oxsym}^{iF}\bigr)=i\,d(F)\oxsym e_{\oxsym}^{iF}\word{for both} d:=\dl^{(0)}_\th(\al)
\word{and}d:=\dl^{(1)}_\th(\al).
$$
In addition, looking at \eqref{eq:th-mu}--\eqref{eq:dl_Q}, we see that
\be\label{eq:dl+d}
\dl_{\al Q}:=\int dy\,\,\al(y)\,\dl_{Q(y)}=
-i\bigl(\dl^{(0)}_\th(\al)+\dl^{(1)}_\th(\al)\bigr).
\ee
With these tools we can give a more concise equivalent reformulation 
of the on-shell MWI \eqref{eq:MWI}:
\begin{align}\label{eq:MWI-F}
T\Bigl((\del j)(\al)\oxsym e_{\oxsym}^{iF}\Bigr)_0=&
-T\Bigl(\dl_{\al Q}\,F\oxsym e_{\oxsym}^{iF}\Bigr)_0\\
\equiv &i\,T\Bigl(\dl^{(0)}_\th(\al)(F)\oxsym e_{\oxsym}^{iF}\Bigr)_0
+i\,T\Bigl(\dl^{(1)}_\th(\al)(F)\oxsym e_{\oxsym}^{iF}\Bigr)_0,
\quad\forall\al\in\sD(\bM,\bR),\nonumber
\end{align}
which we understand as an identity for formal power series in $F\in\sF_\loc$.

\paragraph{Conservation of the interacting current.} 
As an application of the version \eqref{eq:MWI-F} of the MWI, 
we study current conservation. For $S,G\in\sF_\loc$ let 
\be\label{eq:int-field}
G_{S,0}:=\nS(S)_0^{\star -1}\star T\bigl(e_{\oxsym}^{iS}\oxsym G\bigr)_0
\ee
be the interacting field to the interaction $S$ and corresponding to $G$,
as defined by Bogoliubov \cite{BS59}. 
More precisely, $G_{S,0}$ is a formal power series in $S$ and 
to zeroth order in $S$ it agrees with $G_0$. For $S$ being the
interaction of scalar QED \eqref{eq:interaction}, we obtain
$$
\dl^{(0)}_\th(\al)(S)=0\word{and} 
\dl^{(1)}_\th(\al)(S)=-2ie\,(\phi\phi^* A^\mu)(g\del_\mu\al).
$$
Now, in the MWI \eqref{eq:MWI-F} we set $F:=S$ and multiply 
with $\nS(S)_0^{\star -1}\star \cdots$. This yields
$$
-j(\del\al)_{S,0}=2e\,(\phi\phi^* A)(g\del\al)_{S,0}.
$$
Omitting the arbitrary testfunction $\al$, this can be written as
conservation of the interacting electromagnetic current:
\begin{align}\label{eq:curr-cons}
&\del^x_\mu J^\mu_S(x)_0=0\word{where}\\
&J^\mu(x):=j^\mu(x)+2eg(x)\,(\phi\phi^* A)(x)=
i\bigl(\phi(x)\,(D^\mu\phi)^*(x)-\phi^*(x)\,D^\mu\phi(x)\bigr)\nonumber
\end{align}
with the covariant derivative $D^\mu_x:=\del^\mu_x+ieg(x)\,A^\mu(x)$.
$J^\mu$ is the Noether current belonging to the invariance of the
total action 
$$
S_0+S=\int dx\,\,\bigl((D^\mu\phi)^*(x)\,D^\mu\phi(x)-
m^2\,\phi^*(x)\phi(x)\bigr)+S_0(A^\mu)
$$
(cf.~\eqref{eq:S0}) under the same global $U(1)$-transformation 
$\phi(x)\to e^{i\al}\,\phi(x)$ as in the preceding sections. 
We recognize a further significant difference to spinor QED: 
the Noether currents $j$ and $J$ belonging to the free and 
interacting theory, respectively, are different.

\subsection{The unitary MWI}\label{sec:unitary-MWI}
The Buchholz--Fredenhagen quantum algebra (``BF-algebra'') \cite{BF19} 
is an abstract C*-algebra (more precisely: a local net of
C*-algebras) which,
given the field content and a classical relativistic Lagrangian,
encodes the pertinent interactions in QFT.
In this generality, the most adequate formulation of symmetries
is an open problem. A concrete algebra $\sA$ fulfilling the
defining relations of the BF-algebra
belonging to the field content of scalar QED and the Lagrangian 
$$
L_0:=(\del\phi^*\del\phi-m^2\,\phi^*\phi)-\tfrac{1}4\,F^{\mu\nu}F_{\mu\nu}
$$
(where $F^{\mu\nu}:=\del^\mu A^\nu-\del^\nu A^\mu$) is given by the
perturbative on-shell $S$-matrices \eqref{eq:S-matrix-1}, that is,%
\footnote{By ``$\bigveestar$'' we mean the algebra, under the 
star product, generated by members of the indicated set. Also the 
analogous algebra generated by the \emph{off-shell} $S$-matrices
(i.e., without restriction to $\sC_{S_0}$) fits into the definition
of the BF-algebra for the same field content and the same Lagrangian $L_0$; 
however, in view of the MWI, we prefer in the following to work on-shell.}
$$
\sA:=\bigveestar \set{\nS(F)_0\,\big\vert\,F\in\sF_\loc}.
$$
For this algebra, the above mentioned problem amounts to the
task of finding an equivalent reformulation of the MWI in terms of 
the maps $\sF_\loc\ni F\to \nS(F)_0$; in contrast to \eqref{eq:MWI-F},
expressions of the type $T\bigl(G\oxsym e_{\oxsym}^{iF}\bigr)_0$ must not appear.

For scalar QED, Fredenhagen has noted that the following conjectured identity 
\cite{BDFR20} (see also \cite{R20}) would serve the purpose, which has some 
analogy to the Schwinger-Dyson equation: let
\be\label{eq:trafo}
\phi_\al(x):=\phi(x)\,e^{i\al(x)},\quad \phi^*_\al(x):=\phi^*(x)\,e^{-i\al(x)},\quad
F_\al:=F(\phi_\al,\phi^*_\al,A^\mu),
\ee
where $\al\in\sD(\bM,\bR)$ (see Remark \ref{rmk:notgauge} below)
and define
\be\label{eq:dlL0(al)}
\dl L_0(\al):=\int dx\,\,\bigl(L_0(x)_\al-L_0(x)\bigr). 
\ee
On the r.h.s.~the range of integration is only $\supp\al$,
that is, a bounded region. The conjecture asserts that the time-ordered 
product can be renormalized such that
\be\label{eq:conject}
\nS\bigl(F_\al+\dl L_0(\al)\bigr)_0=\nS(F)_0,\quad\forall F\in\sF_\loc,\,\,
\al\in\sD(\bM,\bR).
\ee
We understand \eqref{eq:conject} as identity
for \emph{formal power series in $\hbar,F$ and $\alpha$},%
\footnote{The dependence on $\hbar$ is not visible in our notations since we
have set $\hbar:=1$, to simplify the notations.}
and we will call it the ``unitary MWI'', because it expresses the MWI in an 
equivalent way (as we show below) in terms of the $S$-matrix.

Setting $F:=0$ the unitary MWI reduces to
\be
\nS\bigl(\dl L_0(\al)\bigr)_0=1.
\ee
For illustration we explicitly compute $\dl L_0(\al)$.
Taking into account that
\be\label{eq:dx(phial)}
\del_x\phi_\al(x)=(\del\phi)(x)\,e^{i\al(x)}+i\,\phi_\al(x)\,\del\al(x)
\ee
and the analogous relation for $\del_x\phi^*_\al(x)$, we obtain
\be\label{eq:dlL_0(al)}
\dl L_0(\al)=-(\del j)(\al)+(\phi^*\phi)\bigl((\del\al)^2\bigr).
\ee
The following Theorem supports the conjecture:

\begin{thm}\label{th:unitary-MWI}
The unitary MWI \eqref{eq:conject} is equivalent to the 
on-shell MWI \eqref{eq:MWI-F}, when the latter is interpreted as 
an identity which should hold for \emph{all} $F\in\sF_\loc$ and
\emph{all} $\al\in\sD(\bM,\bR)$.
\end{thm}

\begin{remk}\label{rmk:notgauge}
Before giving the proof, we point out that $\alpha$ in \eqref{eq:trafo} having 
compact support does \emph{not} mean that the transformation underlying 
the unitary MWI is a local gauge transformation. Specifically, 
$A^\mu$ is not transformed. The test function $\alpha$ is used to control the
dependence of functionals on the scalar field only, and its localization 
means that the transformation acts non-trivially only in a bounded 
region. Indeed, the Theorem does not hold true for local gauge transformations, 
in particular the relation \eqref{eq:Fa} becomes wrong, because then 
the l.h.s.\ of \eqref{eq:Fa} contains additional terms coming from the transformation 
of $A^\mu$, but the r.h.s.\ does not contain such terms.
A second reason becomes apparent by looking at the model containing only the 
electromagnetic field and assuming that $\al$ is a local gauge transformation. 
Then, it holds that $\dl L_0(\al)=0$. Hence, the conjectured formula 
\eqref{eq:conject} would be trivial for all observables $F$ (i.e.,
$F_\al=F$), hence worthless.
\end{remk}

\begin{proof}
Let $0\not= \bt\in\sD(\bM,\bR)$ be arbitrary and let $\al(x):=a\,\bt(x)$
with $a\in\bR$. To prove that the MWI \eqref{eq:MWI-F} implies the unitary 
MWI \eqref{eq:conject}, let $\bt$ be fixed and interpret the l.h.s.\ of
\eqref{eq:conject} as a function $f(a)$ of $a\in\bR$, explicitly
$$
f(a):=T\bigl(e_{\oxsym}^{iG(a)}\bigr)_0,\word{with} G(a):=F_{a\bt}+\dl L_0(a\bt).
$$
Since this function is differentiable (as we see from the explicit formulas) and since 
the unitary MWI holds trivially true for $a=0$, it suffices to show that 
$\frac{d}{da}f(a)=0$ for all $a\in\bR$ --
as a consequence of the MWI
\eqref{eq:MWI-F}.\footnote{For this function $f$, 
differentiability and the application of the fundamental theorem of calculus 
are understood in the sense of formal power series and functionals, that is, 
for each term of the formal series and applied to any fixed classical configuration.}

To prove $\frac{d}{da}f(a)=0$, we use the following crucial relation for $F_{a\bt}$:
\be\label{eq:Fa}
\frac{d}{da}F_{a\bt}=-\dl_{\bt Q}\,F_{a\bt}\quad\forall F\in\sF_\loc.
\ee
\emph{Proof of \eqref{eq:Fa}}: we compute the l.h.s.\ by using 
$\frac{d}{da}\phi_\al(x)=i\bt(x)\,\phi_\al(x)$:
\begin{align*}
\frac{d}{da}F_{a\bt}=&\int dy\,\Bigl(
\frac{d\phi_\al(y)}{da}\,\frac{\dl F_{a\bt}}{\dl\phi_\al(y)}
+\frac{d\phi^*_\al(y)}{da}\,\frac{\dl F_{a\bt}}{\dl\phi^*_\al(y)}\Bigr)\\
=&i\int dy\,\,\bt(y)\Bigl(
\phi_\al(y)\,\frac{\dl F_{a\bt}}{\dl\phi_\al(y)}
-\phi^*_\al(y)\,\frac{\dl F_{a\bt}}{\dl\phi^*_\al(y)}\Bigr).
\end{align*}
Taking into account that $\frac{\dl\phi_\al(z)}{\dl\phi(y)}=\dl(z-y)\,e^{i\al(z)}$,
which implies
\be
\phi(y)\,\frac{\dl}{\dl\phi(y)}=
\int dz\,\,\phi(y)\,\frac{\dl\phi_\al(z)}{\dl\phi(y)}\frac{\dl}{\dl\phi_\al(z)}
=\phi_\al(y)\,\frac{\dl}{\dl\phi_\al(y)},
\ee
and inserting $Q(y)=-i\phi(y)$ we obtain the assertion \eqref{eq:Fa}:
$$
\frac{d}{da}F_{a\bt}=i\int dy\,\,\bt(y)\Bigl(
\phi(y)\,\frac{\dl F_{a\bt}}{\dl\phi(y)}
-\phi^*(y)\,\frac{\dl F_{a\bt}}{\dl\phi^*(y)}\Bigr)=-\dl_{\bt Q}F_{a\bt}.
$$
This concludes the proof of \eqref{eq:Fa}.

Now let $f\in\sD(\bM,\bR)$ with $f\big\vert_{\supp\bt}=1$. 
Thanks to this property of $f$, it holds that
\be\label{eq:L0(f)}
\dl_{\bt Q}\,L_0(f)=\dl_{\bt Q}\,S_0\word{and}
L_0(f)_{a\bt}-L_0(f)\overset{\eqref{eq:dlL0(al)}}{=}\dl L_0(a\bt)
\ee
where $L_0(f)_{a\bt}$ as a function of $a$ is defined similarly
to $F_{a\bt}=F_\al$, i.e., $\alpha(x)=a\beta(x)$. Note in particular, 
that for both equations in \eqref{eq:L0(f)} also the l.h.s.'s do not depend 
on the choice of $f$.

By applying the relation \eqref{eq:Fa} to $L_0(f)_{a\bt}$, we obtain
\begin{align}\label{eq:dL0/da}
\frac{d\,\dl L_0(a\bt)}{da}\overset{\eqref{eq:L0(f)}}{=}&
\frac{d\,L_0(f)_{a\bt}}{da}\overset{\eqref{eq:Fa}}{=}-\dl_{\bt Q}\,L_0(f)_{a\bt}
\overset{\eqref{eq:L0(f)}}{=}-\dl_{\bt Q} \bigl(\dl L_0(a\bt)\bigr)
-\dl_{\bt Q}\bigl(L_0(f)\bigr)
\nonumber\\
\overset{\eqref{eq:L0(f)}}{=}&-\dl_{\bt Q}\bigl(\dl L_0(a\bt)\bigr)-
\dl_{\bt Q}\,S_0.
\end{align}

Equipped with these tools we are able to verify the vanishing of $\frac{d}{da}f(a)$ for all $a\in\bR$.
The derivative can easily be computed:
\begin{align}\label{eq:dT/da}
\frac{d}{da}f(a)&=i\,T\Bigl(e_{\oxsym}^{iG(a)}\oxsym
\Bigl[\frac{dF_{a\bt}}{da}+\frac{d\,\dl L_0(a\bt)}{da}\Bigr]\Bigr)_0\nonumber\\
&\overset{\eqref{eq:Fa},\eqref{eq:dL0/da}}{=}
-i\,T\Bigl(e_{\oxsym}^{iG(a)}\oxsym
\bigl[\dl_{\bt Q}\,F_{a\bt}+\dl_{\bt Q}\bigl(\dl L_0(a\bt)\bigr)+\dl_{\bt Q}\,S_0\bigr]
\Bigr)_0\nonumber\\
=&-i\,T\Bigl(e_{\oxsym}^{iG(a)}\oxsym
\bigl[\dl_{\bt Q}\,G(a)+\dl_{\bt Q}\,S_0\bigr]\Bigr)_0;
\end{align}
the r.h.s.\ vanishes due to the MWI \eqref{eq:MWI-F} for $G(a)\in\sF_\loc$,
by remembering that $\dl_{\bt Q}\,S_0=\del j(\bt)$ \eqref{eq:dl_Q}.

That the unitary MWI implies the MWI is obvious from our procedure: the
former yields $\frac{d}{da}\big\vert_{a=0} T\bigl(e_{\oxsym}^{iG(a)}\bigr)_0=0$; 
after use of \eqref{eq:dT/da} this is the MWI 
\eqref{eq:MWI-F} for $G(0)=F$ and $\bt$, which is the MWI in its full 
generality, because $F\in\sF_\loc$ and $\bt\in\sD(\bM,\bR)$ are arbitrary.  
\end{proof}

The fact that in this proof model-specific information is used only in 
the verification of the relation \eqref{eq:Fa}, indicates that the 
conjecture is valid also for other models with other symmetry 
transformations of the basic fields -- see \cite{BDFR20}.

\section{Proof of the Master Ward Identity}
\label{sec:MWI-proof}
 
In this section we prove the first main result of this paper, to wit,
that the MWI \eqref{eq:MWI} can be satisfied by a finite renormalization 
of the $T$-product for all $B_1,\ldots,B_n\in \sP_0$, the set of
$$
L:=e\,jA+ge^2\,A^2\phi^*\phi\,,\,\,\,j^\mu\word{and all sub\emph{monomials}\/
of these two field polynomials,}
$$
except $e\,jA$ and $ge^2\,A^2\phi^*\phi$ and the individual parts of
$j^\mu$ (i.e.~$i\phi\del^\mu\phi^*$ and $-i\phi^*\del^\mu\phi$)
separately. These exceptions are made to shorten the proof
\footnote{There is no reason to believe that the MWI \eqref{eq:MWI}
  holds not true for a larger version of $\sP_0$ containing these
  exceptions.}; they are justified by the fact that physically relevant are only
$L$ and $j^\mu$, that is, only the \emph{sums} of their individual parts.
The validity of the MWI \eqref{eq:MWI} for the set $\sP_0$ is sufficient for 
conservation of the interacting current \eqref{eq:curr-cons} and (most probably)
for a local construction of the main observables of scalar QED
(including the interacting fields $\phi$ and $\phi^*$), in
analogy to the construction given in 
\cite{DF99} or \cite[Chap.~5]{D19}.

Note here a technical artifice: in the above definition of $L$ we have multiplied 
$e^2$ with $g(x)$, where $g\in\sD(\bM)$ is arbitrary, in order that $L(g)$ agrees with the interaction
$S$ given in \eqref{eq:interaction}.%
\footnote{Hence, as long as we do not perform the adiabatic limit $g(x)\to 1$, we investigate a theory 
with a spacetime dependent coupling constant $eg(x)$.} 
However, in all other elements of $\sP_0$ (in particular, the proper
submonomials of $ge^2\,A^2\phi^*\phi$), we do not include this factor $g$.
Note that all $B_j\in\sP_0$ are 
eigenvectors of $\th$; we will use the notation $b_jB_j:=\th B_j$.

We proceed in analogy with the proof of the QED-MWI in 
\cite[Chap.~5.2.2]{D19}, which relies on \cite[App.~B]{DF99}, 
and in addition
we use specific arguments for scalar QED given in \cite{P20}.
Starting with a $T$-product fulfilling all other renormalization conditions
(including the AWI and the Field Equation)
and proceeding by induction on $n$, the anomalous term (i.e.,
  the possible violation of the MWI) is given by
\begin{align}\label{eq:MWI-anomaly}
(-i)^n\,\Dl^n\bigl(B_1(x_1)&,\ldots,B_n(x_n);y\bigr)_0:=
-\del_y^\mu\,T_{n+1}\bigl(B_1(x_1)\ox\cdots\ox  B_n(x_n)\ox
j_\mu(y)\bigr)_0\nonumber\\
&+\sum_{l=1}^n\dl(y-x_l)\,T_n\bigl( B_1(x_1)\ox \cdots
\ox(\th B_l)(x_l)\ox\cdots\ox  B_n(x_n)\bigr)_0
\nonumber\\
&-\del^\mu_y\Bigl(\sum_{l=1}^n\dl(y-x_l)\,T_n\bigl( B_1(x_1)\ox \cdots
\ox(\th_\mu B_l)(x_l)\ox\cdots\ox  B_n(x_n)\bigr)_0\Bigr).
\end{align}
By using causal factorization of the $T$-products and the validity 
of the MWI to lower orders, one proves that
\be\label{eq:supp(Delta)}
\supp\Dl^n\bigl(B_1(x_1),\ldots,B_n(x_n);y\bigr)_0\subseteq\Dl_{n+1},
\ee
where $\Dl_{n+1}$ denotes the thin diagonal in $\bM^{n+1}$ \eqref{eq:diagonal},
for details see \cite[Chap.~4.2.2]{D19}. Therefore, the MWI
\eqref{eq:MWI} is indeed a renormalization condition, to be
  imposed on the
  definition of $T_{n+1}$ in the next perturbative order.
The task is to remove $\Dl^n\bigl(B_1(x_1),\ldots;y\bigr)_0$ by a finite,
admissible renormalization of 
$T_{n+1}\bigl(B_1(x_1)\ox\cdots\ox  B_n(x_n)\ox j_\mu(y)\bigr)_0$, cf.\ \eqref{eq:nonunique}. By 
``admissible'' we mean that the basic axioms and the above mentioned 
renormalization conditions (i)--(viii) and AWI and FE are maintained. 

The idea of proof goes as follows: in the first two steps we prove
$\int dy\,\,\Dl^n\bigl(\cdots;y\bigr)_0=0$. Because $\Dl^n\bigl(\cdots;y\bigr)_0$ satisfies a version of the causal Wick
expansion, it suffices to study the vacuum expectation value $d_n(B_1,\dots)(x_1-y,\dots):=\om_0\bigl(
\Dl^n( B_1(x_1),\ldots;y)\bigr)$. By \eqref{eq:supp(Delta)}, $d_n(B_1,\dots)(x_1-y,\dots)$ 
is a linear combination of derivatives
of $\dl(x_1-y,\ldots,x_n-y)$. By a version of the Poincar\'e Lemma, the result of the first two steps implies that
$d_n(B_1,\dots,)(x_1-y,\dots)=\del^y_\mu
u^\mu_n(B_1,\dots)(x_1-y,\dots)$ for some numerical distributions
$u^\mu_n(B_1,\dots)(x_1-y,\dots)$ supported also on the thin diagonal.
Therefore, the finite renormalization
$t_{n+1}(B_1,\dots,j^\mu)\,\to\,t_{n+1}(B_1,\dots,j^\mu)+(-i)^n\,u^\mu_n(B_1,\dots)$
removes $\Dl^n\bigl(\cdots;y\bigr)_0$. One has to verify that the
  other renormalization conditions are maintained. Only one instance
  poses a serious difficulty: if at least one of the $B_j$'s is a current $j^\nu$,
it is not clear that this finite renormalization maintains the invariance of $t_{n+1}(B_1,\dots,j^\mu)$ under permutation of 
the $j$'s. We solve this problem by proceeding case by case.

\textbf{Step 1}: Similarly to \cite[Exer.~5.1.7]{D19} one shows that
\be\label{eq:Q-phi}
[Q^\phi, B(x)_0]_\star =(\th B)(x)_0,\word{with}
Q^\phi:=\int d\vec y\,\,j^0(t,\vec y)_0,
\ee
where the time $t\in\bR$ is arbitrary and $[\,\cdot\,,\,\cdot\,]_\star$ denotes
the commutator w.r.t.~the star product. The integral
    $Q^\phi$ in \eqref{eq:Q-phi} is meant symbolically, but its commutator  understood as the
    integral over $[j^0(t,\vec y)_0,B(x)_0]_\star$ is well-defined by locality.

As it becomes clear below in Step 2,
a necessary condition for the asserted MWI \eqref{eq:MWI}
is \emph{charge number conservation}, which is a generalization
of the relation \eqref{eq:Q-phi} to time-ordered products of order $n\geq 2$, 
explicitly:
\begin{align}\label{eq:cn-cons}
[Q^\phi, T_n\bigl(B_1(x_1)\oxyox B_n(x_n)\bigr)_0]_\star
=&\sum_{l=1}^n\,T_n\bigl(B_1(x_1)\oxyox(\th B_l)(x_l)\oxyox B_n(x_n)\bigr)_0
\nonumber\\
=&T_n\bigl(B_1(x_1)\oxyox B_n(x_n)\bigr)_0\cdot\sum_{l=1}^nb_l.
\end{align}
To explain how one can satisfy this relation, we first study a necessary
condition for it, which is obtained by taking the VEV of \eqref{eq:cn-cons}: 
using $\om_0\bigl([Q^\phi, F]_\star\bigr)=0$ for any $F\in\sF$ we get
\be\label{eq:nc-charge}
t_n(B_1,\ldots,B_n)=0 \word{if} \sum_{l=1}^nb_l\not= 0\ .
\ee
In the inductive construction of the $T$-products, the property 
\eqref{eq:nc-charge} can get lost only in the extension to the thin diagonal,
i.e., \eqref{eq:nc-charge} is a renormalization condition. To fulfill it, we
simply extend zero by zero --- this is compatible with all other 
renormalization conditions. That \eqref{eq:nc-charge} is also sufficient for
\eqref{eq:cn-cons} can be verified by means of the causal Wick expansion -- for details see
\cite[Chap.~5.2.2]{D19}. So we assume in the following steps, that the $T$-products satisfy also 
\eqref{eq:cn-cons}.
We work here with the charge number operator $\th$ only; 
$\th_\mu$ does not play any role here.

\textbf{Step 2}: In this step we prove that
charge number conservation \eqref{eq:cn-cons} is equivalent to the relation
\be\label{eq:int-anomaly}
\int dy\,\,\Dl^n\bigl(B_1(x_1),\ldots,B_n(x_n);y\bigr)_0=0\ .
\ee 
Here, we understand the l.h.s.~as a distribution in $(x_1,\dots,x_n)$, i.e., 
smeared out with an arbitrary $h(x_1,\dots,x_n)\in\sD(\bM^n)$.
Thanks to this and  \eqref{eq:supp(Delta)}, the integral is well defined.

For a given configuration $(x_1,...,x_n)\in\bM^n$ let $\sO\subset \bM$
be an open double cone (i.e., the nonempty intersection of an
  open forward lightcone with an open backward lightcone) with $x_1,\ldots ,x_n\in\sO$; in addition let $f$ be
an arbitrary test function satifying $f\vert_{\ovl \sO}=1$.
Thanks to \eqref{eq:supp(Delta)}, we may write
\begin{align}\label{eq:int-anomaly-1}
(-i)^n\int dy\,\,\Dl^n\bigl(&B_1(x_1),\ldots,B_n(x_n);y\bigr)_0=
(-i)^n\,\int dy\,\,f(y)\,\Dl^n\bigl(B_1(x_1),\ldots,B_n(x_n);y\bigr)_0
\nonumber\\
=&-\int dy\,\,f(y)\,\del_y^\mu\,T\bigl(B_1(x_1)\ox\cdots\ox 
B_n(x_n)\ox j_\mu(y)\bigr)_0\nonumber\\
&+\sum_{l=1}^n T\bigl(B_1(x_1)\ox \cdots
\ox(\th B_l)(x_l)\ox\cdots\ox B_n(x_n)\bigr)_0\nonumber\\
&+\sum_{l=1}^n \del^\mu f(x_l)\,T\bigl( B_1(x_1)\ox \cdots
\ox(\th_\mu B_l)(x_l)\ox\cdots\ox  B_n(x_n)\bigr)_0,
\end{align}
where we have integrated out the $\dl$-distributions. Compared with
\cite[eqn.~(5.2.20)]{D19}, there is an additional term appearing in the last 
line. However, because $f\vert_{\ovl \sO}=1$, this term vanishes. So we may 
continue as in that reference: we decompose $\del^{\mu} f=a^{\mu}-b^{\mu}$
such that $\supp a^{\mu}\cap (\sO+\overline{V}_-)=\emptyset$ and  
$\supp b^{\mu}\cap (\sO+\overline{V}_+)=\emptyset$. By causal
factorization of the $T$-products, the first term on the r.h.s.~of 
\eqref{eq:int-anomaly-1} becomes 
\begin{align*}
&j^{\mu}(a_{\mu})_0\star T\bigl(B_1(x_1)\oxyox B_n(x_n)\bigr)_0-
T\bigl(B_1(x_1)\oxyox B_n(x_n)\bigr)_0\star j^{\mu}(b_{\mu})_0
\nonumber\\
&=[j^{\mu}(a_{\mu})_0\,,\,T\bigl(B_1(x_1)\oxyox B_n(x_n)\bigr)_0]_\star
+T\bigl(B_1(x_1)\oxyox B_n(x_n)\bigr)_0\star j^{\mu}(\partial_{\mu}f)_0.
\end{align*}
The second term vanishes because $\del_\mu j^{\mu}_0=0$. From the Field
Independence of $T$ we know that
$\,\supp T\bigl(\tilde B_1(x_1)\oxyox\tilde B_n(x_n)\bigr)_0\subset{\cal O}$;
therefore, we may vary $a^\mu$ in the set
$$
\set{z\in\bM\,|\,(z-x)^2<0\,\,\,\forall x\in\sO}\word{without affecting}
[j^{\mu}(a_{\mu})_0,T\bigl(\tilde B_1(x_1)\ox\cdots\bigr)_0]_\star\ .
$$
In particular we may choose for $a_\mu$ a smooth approximation
to $\del_\mu \th(c-x^0)=-\dl_{\mu 0}\,\dl(x^0-c)$, 
where $c\in \bR$ is a sufficiently large constant:
$$
a_\mu(x)=-\dl_{\mu 0}\,h(x^0)\word{with} \int dx^0\,\, h(x^0)=1\ ,\quad
h\in\sD([c-\eps,c+\eps])
$$ 
for some $\eps >0$. Then, we obtain
\begin{align*}
[j^{\mu}(a_{\mu})_0\,,&\,T(\ldots)_0]_\star=-\int dx^0\,\,h(x^0)\int d\vec x\,\,
[j^0(x^0,\vec x)_0\,,\,T(\ldots)_0]_\star\\
=&-\int d\vec x\,\,[j^0(c,\vec x)_0\,,\,T(\ldots)_0]_\star\,\Bigl(
\int dx^0\,\,h(x^0)\Bigr)=-[Q^\phi\,,\,T(\ldots)_0]_\star\ .
\end{align*}
Inserting these results into \eqref{eq:int-anomaly-1}, we see that charge number conservation
\eqref{eq:cn-cons} implies the assertion \eqref{eq:int-anomaly}.

\textbf{Step 3}: Following the proof of the QED-MWI we list some 
structural properties of the anomalous term $\Dl^n(\cdots)$ defined in 
\eqref{eq:MWI-anomaly}, for the validity of these properties see the 
above mentioned references.

First, $\Dl^n(\cdots)$ satisfies the following version of the causal Wick expansion (cf.~\eqref{eq:causal-Wick-expan}):
\be\label{eq:anomaly-caus-Wick}
\Dl^n\bigl( B_1(x_1),\ldots, B_n(x_n);y\bigr)=
\sum_{\unl B_l \subseteq B_l}\!  
d_n(\unl B_1,\dots, \unl B_n)(x_1-y,\dots)\,
\ovl B_1(x_1)\cdots  \ovl B_n(x_n)\ ,
\ee
where the sum runs over all submonomials $\unl B_l$ of $B_l$
(where $1\leq l\leq n$), and 
\be\label{eq:om(Dl)}
d_n(B_1,\dots, B_n)(x_1-y,\dots,x_n-y):=\om_0\Bigl(
\Dl^n\bigl( B_1(x_1),\ldots, B_n(x_n);y\bigr)\Bigr)
\in\sD'(\bM^{n}).
\ee

If $B_j=L$ and $\unl B_j$ is a proper submonomial of $ge^2\,A^2\phi^*\phi$, we include the factor $g(x_j)$ in $\ovl B_j(x_j)$. 
This is consistent, since, due to $T(\cdots\ox g(x_j)(A^2\phi^*\phi)(x_j)\ox\cdots)=g(x_j)\,T(\cdots\ox (A^2\phi^*\phi)(x_j)\ox\cdots)$, 
it holds that 
$\Dl^n\bigl(\ldots,g(x_j) (A^2\phi^*\phi)(x_j),\ldots;y\bigr)=g(x_j)\,\Dl^n\bigl(\ldots, (A^2\phi^*\phi)(x_j),\ldots;y\bigr)$.

The validity of \eqref{eq:anomaly-caus-Wick} can be
    traced back to the validity of the
causal Wick expansion for the $T$-products appearing on the r.h.s.~of \eqref{eq:MWI-anomaly}; the fact that there is no term
in \eqref{eq:anomaly-caus-Wick} coming from proper submonomials of $\del^\mu j_\mu$ is due to the validity of the axiom
FE -- for details see   \cite[Thm.~4.3.1 and Chap.~5.2.2]{D19}.

If one of the $B_j$'s is linear in the basic fields, e.g., $B_1=\del^a\phi$
with $a\in\bN^4$, the axiom FE determines \emph{uniquely} 
$T_{n+1}\bigl(\del^a\phi(x_1)\oxyox j_\mu(y)\bigr)$ in
  terms of $T_k$ with $k\leq n$, that is, there 
is no freedom to remove $\Dl^n\bigl(\del^a\phi(x_1)\ox\cdots\bigr)$
by a finite renormalization of this $T$-product. However, as verified in
\cite[Exer.~4.3.3]{D19}, the validity of the axiom FE implies
that $\Dl^n(\cdots)$ vanishes in this case.
Therefore, on the r.h.s.\ of \eqref{eq:anomaly-caus-Wick}, the sum is 
restricted to submonomials  $\unl B_l$ of $B_l$ which are at least quadratic in 
the basic fields for all $l$.

From \eqref{eq:supp(Delta)} we conclude that $d_n( B_1,\dots, B_n)(x_1-y,\dots,x_n-y)$ 
is a linear combination of derivatives
of $\dl(x_1-y,\ldots,x_n-y)$. Using in addition a version of the Poincar\'e
Lemma (more precisely, \cite[Lemma 4.5.1]{D19}), the property 
$$
\int dy\,\,d_n(B_1,\dots, B_n)(x_1-y,\dots,x_n-y)=0
$$ 
(which is obtained by taking the VEV of the corresponding relation 
for $\Dl^n$ \eqref{eq:int-anomaly}) implies that we can write
$d(B_1,\dots,B_n)$ as
\be\label{eq:d=div(u)}
d_n(B_1,\dots,B_n)(x_1-y,\dots,x_n-y)=\del^y_\mu 
u^\mu_n(B_1,\dots,B_n)(x_1-y,\dots,x_n-y)
\ee
where $u^\mu_n(B_1,\dots,B_n)$ is Lorentz covariant and of the form
\be\label{eq:u}
u^\mu_n(B_1,\dots,B_n)(x_1-y,\dots)=\sum_{a\in\bN^{4n}}
 C_a(B_1,\ldots,B_n)\,\del^a\dl(x_1-y,\ldots).
\ee
Since $d_n(B_1,\dots,B_n)$ is defined by the VEV of
the r.h.s.\ of \eqref{eq:MWI-anomaly},%
\footnote{See \eqref{eq:tn} for the definition of $t_{n+1}$ and $t_n$, 
respectively.}
\begin{align}\label{eq:MWI-d}
(-i)^n\,d_n(&B_1,\dots,B_n)(x_1-y,\dots,x_n-y):=
-\del_\mu^y\,t_{n+1}(B_1,\dots,B_n,j^\mu)(x_1-y,\dots,x_n-y)\nonumber\\
&+\sum_{l=1}^n b_l\,\dl(y-x_l)\,t_n(B_1,\dots,B_n)
(x_1-x_n,\ldots,x_{n-1}-x_n)\nonumber\\
&-\del_\mu^y\Bigl(\sum_{l=1}^n\dl(y-x_l)\,t_n(B_1,\dots,(\th^\mu B_l),\dots,B_n)
(x_1-x_n,\ldots,x_{n-1}-x_n)\Bigr),
\end{align}
we obtain an upper bound for the
scaling degree of $d_n(B_1,\dots,B_n)$ by the maximum 
of the scaling degrees of the terms standing on the 
r.h.s.\ of \eqref{eq:MWI-d}. Proceeding this way, we see
that the sum over $a$ in \eqref{eq:u} is bounded by
\be\label{eq:bound(a)}
|a|\leq\om(B_1,\ldots,B_n)-1\word{where}
\om(B_1,\ldots,B_n):=\sum_{j=1}^n\dim B_j+4-4n.
\ee

\textbf{Step 4}: Obviously, the finite renormalization
\be\label{eq:remove-anom}
t_{n+1}(B_1,\dots,B_n,j^\mu)\,\,\to\,\,
t_{n+1}(B_1,\dots,B_n,j^\mu)+(-i)^n\,u^\mu_n(B_1,\dots,B_n)
\ee
removes the anomalous term $d_n(B_1,\dots,B_n)$. 
Looking at the causal Wick expansion of $\Dl_n$
\eqref{eq:anomaly-caus-Wick} we conclude:
performing the finite renormalization \eqref{eq:remove-anom} for all 
$B_1,\ldots,B_n\in\sP_0$ being at least quadratic in the basic fields,
the MWI \eqref{eq:MWI} is proved for all $B_1,\ldots,B_n\in\sP_0$,
provided that all these finite renormalizations are admissible, that is,
they maintain the basic axioms and the renormalization conditions (v)-(viii), 
AWI and FE.

This is obvious for%
\footnote{We use the numbering of the axioms given in Appendix \ref{sec:axioms-T}.}
(i) Linearity (the set $\sP_0$ is linearly independent), (iv) Causality, (v) Field Independence, (vii) Poincar\'e Covariance and
FE Field Equation (all $B_j$'s in \eqref{eq:remove-anom} are at least bilinear in the basic fields).
The maintenance of the axiom (viii) Scaling Degree follows from \eqref{eq:bound(a)}. 

Concerning the axiom (vi) $*$-Structure, first note that $d_n(B_1^*,\dots,B_n^*)=\ovl{d_n(B_1,\dots,B_n)}$ 
\cite[formula (5.2.26)]{D19}. Hence, to preserve this axiom, we may replace $u_n(B_1,\dots,B_n)$ by
$\frac{1}2\,\bigl(u_n(B_1,\dots,B_n)+\ovl{u_n(B_1^*,\dots,B_n^*)}\bigr)$.

To maintain the axiom AWI, we also perform a finite renormalization of $t_{n+1}(B_1,\dots,\del B_j,\dots,j^\mu)$
and $(B_1,\dots,B_n,\del^\nu j^\mu)$ such that the renormalized $t_{n+1}$ satisfies the AWI: 
$\del_{x_j}t_{n+1}(\dots, B_j,\dots)=t_{n+1}(\dots,\del B_j,\dots)$. In particular, for $B_j=\phi\phi^*$
we may renormalize $t_{n+1}(\dots,\del B_j,\dots)$ independently from the renormalization of $t_{n+1}(\dots,j,\dots)$, since
$j=i(\phi\del\phi^*-(\del\phi)\phi^*)$. The validity of the AWI for
$t_{n+1}$ implies the validity of the AWI for the pertinent $T_{n+1}$ constructed by the causal Wick expansion, as explained
in \cite[Chap.~3.2.4, Step 3]{D19}.
(The wave front set condition poses no problem by translation
  invariance of $u_n^\mu$, cf.\ \cite[Remark 1.2.6]{D19}.)

Only the maintenance of the axiom (iii) Symmetry poses a
    serious difficulty:
if at least one of the $B_j$'s is a current $j^\nu$,
it is not clear whether the finite renormalization \eqref{eq:remove-anom}
preserves the invariance of
\be\label{eq:t(B,j)}
t_{n+1}(B_1,\ldots,B_l,j^{\nu_1},\ldots,j^{\nu_k},j^\mu)
(x_{11}-y,\ldots,x_{1l}-y,x_{21}-y,\ldots,x_{2k}-y)
\ee
(where $l+k=n$) under the permutations of the entries pertaining to the 
currents, that is, under
$(x_{2r},\nu_r)\leftrightarrow (y,\mu)$ for all $1\leq r\leq k$.
Taking additionally into account that $u^\mu_n(B_1,\dots,B_n)$ is not uniquely 
determined by $d_n(B_1,\dots,B_n)$ (one may add to $u^\mu_n$ 
some $\tilde u^\mu_n$ 
with $\del^y_\mu \tilde u^\mu_n=0$), the remaining task can be formulated as 
follows: for any $B_1,\ldots,B_l\in\sP_0\less\set{j^\nu}$ being at least
quadratic in the basic fields, with $l\leq n-1$, and satisfying
\be\label{eq:om>0}
1\leq\om(B_1,\ldots,B_l,j^{\nu_1},\ldots,j^{\nu_{n-l}})=\sum_{s=1}^l\dim B_s+4-n-3l
\ee
(where $\dim j^\nu=3$ is used), we have to find distributions
$u^\mu_n(B_1,\ldots,B_l,j^{\nu_1},\ldots,j^{\nu_k})$ of the form \eqref{eq:u}
having the same permutation
symmetries and the same Lorentz covariance properties as 
$t_{n+1}(B_1,\ldots,B_l,j^{\nu_1},\ldots,j^{\nu_k},j^\mu)$, and which fulfil 
the equation \eqref{eq:d=div(u)} for 
$d_n(B_1,\ldots,B_l,$ $j^{\nu_1},\ldots,j^{\nu_k})$ given by \eqref{eq:MWI-d}.

Looking at \eqref{eq:d=div(u)}, note that even if the distributions
  $d_n=\del_\mu^yu_n^\mu$ 
are invariant under the action on coordinates and indices of a group
$G$ of permutations $\pi$, and the operation $\del_\mu^y$ commutes
with this action, then $u_n$ does in general
not share this symmetry; but we may redefine it such that it does. In detail, calling $p_\pi$ the relevant representation of $\pi$ and
$p:=\frac{1}{|G|}\sum_{\pi\in G}p_\pi$ the symmetrizer, we may replace 
\be\label{eq:symmetrization}
u_n \word{by} pu_n,\word{because}\del_\mu^y(pu_n^\mu)=p(\del_\mu^yu_n^\mu)=pd_n=d_n.
\ee

For some $n$-tuples $(B_1,\ldots,B_l,j^{\nu_1},\ldots,j^{\nu_{n-l}})$ 
satisfying \eqref{eq:om>0}, we know
that $d_n(B_1,\ldots,B_l,$ $j^{\nu_1},\ldots,j^{\nu_{n-l}})=0$ due to 
\emph{charge number conservation (CNC)} or \emph{Furry's theorem (FT)}. 
In detail:
\begin{itemize}
\item[\textbf{CNC}] By using \eqref{eq:nc-charge}  
$$
(B_1,\ldots,B_l,j^{\nu_1},\ldots,j^{\nu_{n-l}})\word{satisfying}\sum_{s=1}^l b_s\neq 0
$$
all distributions $t_{n+1}(\ldots)$ and $t_n(\ldots)$ appearing on the
r.h.s.\ of \eqref{eq:MWI-d} vanish, hence $d_n(\ldots)=0$.

\item[\textbf{FT}] \emph{Charge conjugation} is a linear operator 
$\bt_C:\sF\to\sF$ which is given by the relations 
$$
\bt_C(\del^a\phi(x))=\eta_C\,\del^a\phi^*(x),\quad
\bt_C(\del^a\phi^*(x))=\ovl\eta_C\,\del^a\phi(x)\word{and}
\bt_C(\del^a A^\mu(x))=-(\del^a A^\mu(x)),
$$
where $\eta_C\in\set{z\in\bC\,\big\vert\,|z|=1}$ is a fixed number,
and by 
$$\bt_C\big\langle f_{p,n,l}^{\mu_1\ldots},(\ox_{i=1}^p 
A_{\mu_i})\ox\phi^{\ox n}\ox(\phi^*)^{\ox l}\big\rangle:=
\big\langle f_{p,n,l}^{\mu_1\ldots},(\ox_{i=1}^p \bt_C A_{\mu_i})\ox(\bt_C\phi)^{\ox n}
\ox(\bt_C\phi^*)^{\ox l}\big\rangle
$$
(where \eqref{eq:F} is used); 
for details about charge conjugation in the star-product formalism of 
this paper see \cite[Chap.~5.1.5]{D19}.

\emph{Charge Conjugation Invariance} is the condition
\be\label{eq:cc-inv}
\bt_C\circ T_n=T_n\circ\bt_C^{\ox n}
\ee
on the $T$-product. Analogously to spinor QED (see, e.g.,
  \cite[Chap.~5.1.5]{D19} again), 
one verifies that this is an additional renormalization 
condition, which can be fulfilled such that all other renormalization 
conditions are preserved.

\emph{Furry's theorem} is a consequence of \eqref{eq:cc-inv}, obtained by 
using that $\om_0\circ\bt_C =\om_0$. It states: 
Let $A_i,B_j\in\sP,\, i = 1,\dots,r,\, j = 1,\dots,s$ with 
$\bt_C A_i = A_i$ and $\bt_C B_j = -B_j$ for all $i,j$. Then it holds that
\be\label{eq:furry}
t_{r+s}(A_1,\dots,A_r,B_1,\dots,B_s) = 0\word{if $s$ is odd.}
\ee

Looking at the definition of $d_n(\ldots)$ \eqref{eq:MWI-d} for the $n$-tuples
$$
(L,\ldots,L,j,j),\quad(L,\ldots,L,A\phi^*\phi,j),
$$ 
we verify that all distributions $t_{n+1}(\ldots)$ and $t_n(\ldots)$ appearing 
on the r.h.s.\ vanish, due to \eqref{eq:furry}, hence $d_n(\ldots)=0$. For this
verification we also use the following relations:
\begin{align*}
&\bt_C L=L,\quad\bt_C j^\mu=-j^\mu,\quad\th L=0,\quad\th j=0,\quad
\th^\mu L=-2ieA^\mu\phi^*\phi,\\
&\bt_C(A^\mu\phi^*\phi)=-A^\mu\phi^*\phi ,\quad\bt_C(\th^\mu j^\nu)=\th^\mu j^\nu,
\quad\th(A\phi^*\phi)=0,\quad\th^\mu(A^\nu\phi^*\phi)=0.
\end{align*}
\end{itemize}

There remain the $n$-tuples listed in the following table, that we shall 
study case-by-case. In each case, the 
distribution $u^\mu_n(\cdots)$
for the finite renormalization in \eqref{eq:remove-anom} stands 
for a Lorentz tensor of rank $\geq 2$ according to the entries
$B_1,\dots,B_n$. It is a multiple of the
total $\delta$-distribution in all arguments in all cases except Case
1, where the scaling degree admits two derivatives. Therefore, we
begin with the simpler cases 3 and 2, before we turn to the more
delicate case 1. We use the labelling of the $x$ variables  (i.e., of the
arguments of $B_1,\ldots,B_n$) indicated in \eqref{eq:t(B,j)}.

\begin{table}[h!]
		\centering
		\renewcommand{\arraystretch}{1.3}	
		\begin{tabular}{l|c|c}
\multicolumn{1}{c|}{$B_1,\dots,B_n$} & $\omega(B_1,\dots,B_n)$ & case number\\
\hline			
$\underbrace{L,\dots,L}_{n-1},j^{\nu}$ & 3 &  1\\
$\underbrace{L,\dots,L}_{n-3},j^{\nu_1},j^{\nu_2},j^{\nu_3}$ & 1 & 2a\\
$\underbrace{L,\dots,L}_{n-3},A^{\nu_1}\phi^*\phi,A^{\nu_2}\phi^*\phi,j^{\nu_3}$ 
& 1 & 2b\\
$\underbrace{L,\dots,L}_{n-3},A^{\nu_1}\phi^*\phi,j^{\nu_2},j^{\nu_3}$ & 1 & 2c\\
$\underbrace{L,\dots,L}_{n-2},\phi^*\phi,j^{\nu}$& 1 & 3 \\
$\underbrace{L,\dots,L}_{n-2},A^2,j^{\nu}$& 1 & 3 \\
$\underbrace{L,\dots,L}_{n-3},A^2\phi,A^2\phi^*,j^{\nu}$& 1 & 3 \\
$\underbrace{L,\dots,L}_{n-3},A\del\phi,A\del\phi^*,j^{\nu}$& 1 & 3 \\
$\underbrace{L,\dots,L}_{n-3},A^2\phi,A\del\phi^*,j^{\nu}$& 1 & 3 \\
$\underbrace{L,\dots,L}_{n-3},A\del\phi,A^2\phi^*,j^{\nu}$& 1 & 3 
		\end{tabular}
\end{table}

\bigskip

\textbf{Cases 3:} By \eqref{eq:d=div(u)}--\eqref{eq:bound(a)}, the
renormalization is a Lorentz tensor $u^{\mu\nu}_n$ of rank $2$. The 
only possibility is
$$
u^{\mu\nu}_n(x_{11}-y,\ldots,x_2-y)=C\,g^{\mu\nu}\,\dl(x_{11}-y,\ldots,x_2-y),
$$
for some $C\in\bC$. 
Obviously, $u^{\mu\nu}_n(\ldots,x_2-y)$ is invariant under 
$(\nu,x_2)\leftrightarrow(\mu,y)$, hence, the finite renormalization
\eqref{eq:remove-anom} is admissible in this case.

\medskip

\textbf{Cases 2a,b,c:} Here, $u^\mu_n$ is a Lorentz tensor of rank $4$ which 
is a multiple of the $\dl$-distribution,
\begin{align}\label{eq:u3}
u^{\mu\nu_1\nu_2\nu_3}_n(&x_{11}-y,\ldots,x_{21}-y,\ldots,x_{23}-y)=\\
&(C_1\,g^{\mu\nu_1}g^{\nu_2\nu_3}+C_2\,g^{\mu\nu_2}g^{\nu_1\nu_3}+
C_3\,g^{\mu\nu_3}g^{\nu_1\nu_2})
\prod_{r=1}^{n-3}\dl(x_{1r}-y)\cdot\prod_{s=1}^3\dl(x_{2s}-y),\nonumber
\end{align}
for some $C_k\in\bC$. The totally antisymmetric tensor 
$\epsilon^{\mu\nu_1\nu_2\nu_3}$ is ruled out because we may redefine $u^{\mu\nu_1\nu_2\nu_3}_n$ such that it is
symmetric under $(\nu_{s_1},x_{2s_1})\leftrightarrow(\nu_{s_2},x_{2s_2})$ for at least one pair $(s_1,s_2)$, see \eqref{eq:symmetrization}. 
Similarly, up to a redefinition of $u_n$, we may assume that $u_n^{\mu\nu_1\nu_2\nu_3}$ shares the symmetry of 
$d_n^{\nu_1\nu_2\nu_3}(B_1,\ldots,j^{\nu_3})$ under permutation(s) of the pairs
\begin{align}\label{eq:permutations}
\begin{cases}(\nu_1,x_{21}),(\nu_2,x_{22}),(\nu_3,x_{23})&\word{in case 2a}\\
(\nu_1,x_{21}),(\nu_2,x_{22})&\word{in case 2b}\\
(\nu_2,x_{22}),(\nu_3,x_{23})&\word{in case 2c}\end{cases};
\end{align}
Consequently
\begin{align}\label{eq:C}
\begin{cases}C_1=C_2=C_3&\word{in case 2a}\\
C_1=C_2&\word{in case 2b}\\
C_2=C_3&\word{in case 2c}\end{cases}.
\end{align}
Hence, in cases 2a and 2b we have accomplished that $u_n^{\mu\nu_1\nu_2\nu_3}$
has the needed permutation symmetries to be an admissible finite
renormalization.

In the case 2c, the symmetry of
  $u_n^{\mu\nu_1\nu_2\nu_3}$ under permutations of the three currents
  requires $C_1=C_2=C_3$, which is not already secured by \eqref{eq:C}.
To complete the proof for case 2c, we claim that
\be\label{eq:dd-2c}
\del_{\nu_3}^{x_{23}} d_n(L,\ldots,L,A^{\nu_1}\phi^*\phi,j^{\nu_2},j^{\nu_3})
(x_{11}-y,\ldots,x_{21}-y,\ldots,x_{23}-y)
\ee
is invariant under $x_{23}\leftrightarrow y$. 
To verify this claim, we insert the definition of $d_n$ \eqref{eq:MWI-d}.
Up to a global prefactor $i^n$ we obtain for $\del_{\nu_3}^{x_{23}} d_n$ the 
following sum of terms:%
\footnote{We work here with a modified notation, which ignores that
$t_{n+1}$ and $t_n$ depend on the relative coordinates only; however, it makes 
the computation more intelligible.}
\begin{align}
&-\del_{\nu_3}^{x_{23}}\del_\mu^y t _{n+1}\bigl(L(x_{11}),\ldots,(A^{\nu_1}\phi^*\phi)
(x_{21}),j^{\nu_2}(x_{22}),j^{\nu_3}(x_{23}),j^\mu(y)\bigr)\label{eq:dd-1}\\
&-\sum_{l=1}^{n-3}\del_\mu^y\dl(y-x_{1l})\,\del_{\nu_3}^{x_{23}}t _n\bigl(L(x_{11}),
\ldots,(\th^\mu L)(x_{1l}),\ldots,(A^{\nu_1}\phi^*\phi)(x_{21}),
j^{\nu_2}(x_{22}),j^{\nu_3}(x_{23})\bigr)\label{eq:dd-2}\\
&+2ie\,\del^{\nu_2}_y\dl(y-x_{22})\,\del_{\nu_3}^{x_{23}}t _n\bigl(L(x_{11}),
\ldots,(A^{\nu_1}\phi^*\phi)(x_{21}),(\phi^*\phi)(x_{22}),j^{\nu_3}(x_{23})\bigr)
\label{eq:dd-3}\\
&+2ie\,\del_\mu^y\del^\mu_{x_{23}}\Bigl(\dl(y-x_{23})\,t _n\bigl(L(x_{11}),
\ldots,(A^{\nu_1}\phi^*\phi)(x_{21}),j^{\nu_2}(x_{22}),(\phi^*\phi)(x_{23})\bigr)
\Bigr).\label{eq:dd-4}
\end{align}
Obviously, the terms \eqref{eq:dd-1} and \eqref{eq:dd-4} are individually 
invariant under $x_{23}\leftrightarrow y$. To show this for the sum of the
remaining terms, we insert the MWI to order $(n-1)$, which holds by induction:
\begin{align}
\text[\eqref{eq:dd-2}]=&\sum_{l\not= k}\del_\mu^y\dl(y-x_{1l})\,\del_{\nu_3}^{x_{23}}
\dl(x_{23}-x_{1k})\label{eq:dd-5}\\
&\qquad\cdot t_{n-1}\bigl(\ldots,(\th^\mu L)(x_{1l}),\ldots,(\th^{\nu_3}L)(x_{1k}),
\ldots,(A^{\nu_1}\phi^*\phi)(x_{21}),j^{\nu_2}(x_{22})\bigr)\nonumber\\
&-2ie\sum_l\del_\mu^y\dl(y-x_{1l})\,\del^{\nu_2}_{x_{23}}\dl(x_{23}-x_{22})
\label{eq:dd-6}\\
&\qquad\cdot 
t_{n-1}\bigl(\ldots,(\th^\mu L)(x_{1l}),\ldots,(A^{\nu_1}\phi^*\phi)(x_{21}),
(\phi^*\phi)(x_{22})\bigr),\nonumber\\
\text[\eqref{eq:dd-3}]=&-2ie\sum_l\del^{\nu_2}_y\dl(y-x_{22})\,\del_{\nu_3}^{x_{23}}
\dl(x_{23}-x_{1l})\label{eq:dd-7}\\
&\qquad\cdot t_{n-1}\bigl(\ldots,(\th^{\nu_3}L)(x_{1l}),\ldots,
(A^{\nu_1}\phi^*\phi)(x_{21}),(\phi^*\phi)(x_{22})\bigr)\nonumber.
\end{align}
We see that \eqref{eq:dd-5} is separately invariant under 
$x_{23}\leftrightarrow y$ and that the sum \eqref{eq:dd-6}$+$\eqref{eq:dd-7}
also has this symmetry. Hence, the asserted symmetry of \eqref{eq:dd-2c} 
holds indeed true. 

So we know that
$$
0=\del_{\nu_3}^{x_{23}}\del_\mu^y u^\mu_n(L,\ldots,A^{\nu_1}\phi^*\phi,j^{\nu_2},j^{\nu_3})
(x_{11}-y,\ldots,x_{21}-y,\ldots,x_{23}-y)-(x_{23}\leftrightarrow y).
$$
Inserting the formula \eqref{eq:u3} for $u^{\mu\nu_1\nu_2\nu_3}_n$ into this 
expression and taking into account that $C_2=C_3$ \eqref{eq:C}, we obtain 
that $C_1=C_2(=C_3)$. Hence, also in the case 2c, $u^{\mu\nu_1\nu_2\nu_3}_n$ is an 
admissible finite renormalization.

\medskip

\textbf{Case 1:} Here, $u_n^{\mu\nu}$ is defined by
\begin{align}\label{eq:case1}
&- \del_{\mu}^y \, t_{n+1}\bigl(\overbrace{L,\dots,L}^{m:=n-1},j^{\nu},j^{\mu} \bigr) 
(x_{11}-y,\dots,x_{1m}-y,x_2-y)  \nonumber\\
&- \sum_{l=1}^m\del_{\mu}^y\dl(y-x_{1l})\,
t_n\bigl(L,\dots,(\th^\mu L),\ldots,L,j^\nu\big) (x_{11}-x_2,\dots,x_{1m}-x_2)
\nonumber\\
&+2i\,\del_y^{\nu} \Bigl(\delta(y-x_2)\,
t_n\bigl(L,\dots,L,\phi^* \phi\bigr)(x_{11}-x_2,\dots,x_{1m}-x_2)\Bigr)\nonumber\\
=:&(-i)^n\, \partial_{\mu}^y u^{\mu\nu}_n(x_{11}-y,\dots,x_{1m}-y,x_2-y),
\end{align}
where $u_n^{\mu\nu}$ is a Lorentz tensor of rank $2$ which is a polynomial
in derivatives of the $\dl$-distribution of order $\leq 2$. Taking again into account \eqref{eq:symmetrization}, the terms
$\sim \eps^{\mu\nu\al\bt}\del_\al^{x_i}\del_\bt^{x_j}$ are ruled out   
by their antisymmetry. Hence, $u_n^{\mu\nu}$ must be of the form 
\begin{align}\label{eq:hio}
&u^{\mu\nu}_n(\ldots) 
	= \Big( g^{\mu\nu}\sum_{i,j} a_{ij}\, \partial_i^{\alpha}\partial_{j\alpha} +  
	\sum_{i,j} b_{ij}\, \partial_i^{\mu}\partial_j^{\nu}
	+ g^{\mu\nu}c_0 \Big)\,
	\delta(x_{11}-y,\dots,x_{1m}-y,x_2-y) \; , \nonumber \\
&\word{for some} a_{ij}, b_{ij}, c_0 \in \bC \word{and with}
	i,j \in \{11,...,1m,x_2\} \; .
\end{align}
We have to show that $u^{\mu\nu}_n$ is invariant under 
$(\nu,x_2)\leftrightarrow(\mu,y)$. Obviously, for the $c_0$-term this holds 
true; hence we may omit this term in the following.
 
Since the l.h.s.\ of \eqref{eq:case1} is invariant under permutations of
$x_{11},\ldots,x_{1m}$, up to a redefinition \eqref{eq:symmetrization}, we may assume
that $u^{\mu\nu}_n$ shares this permutation symmetry. We now write down 
all possible contributions with two derivatives to $u_n^{\mu\nu}$ satisfying 
this symmetry:
\begin{align}
	\label{eq:basis1}
& g^{\mu\nu}\sum_{k} \square_{k} && 
\sum_{k} \partial_k^{\mu}\partial_k^{\nu} && 
 g^{\mu\nu}\sum_{k\neq l} \partial_k^{\alpha}\partial_{l\alpha} && 
\sum_{k\neq l} \partial_k^{\mu}\partial_l^{\nu},\nonumber\\
&g^{\mu\nu}\partial_{2\alpha}\sum_{k} \partial_{k}^{\alpha} && 
\partial_{2}^{\mu}\sum_{k}\partial_k^{\nu} && 
\partial_{2}^{\nu}\sum_{k}\partial_k^{\mu} && 
g^{\mu\nu}\square_{2} && \partial_{2}^{\mu} \partial_{2}^{\nu} \; ,
\end{align}
where $k,l \in \{11,...,1m\}$. One verifies that these $9$ differential operators -- 
each one applied to the $\delta$-distribution in \eqref{eq:hio} -- are 
linearly independent if $m>1$\footnote{In the simpler case $m=1$,
  two of them are zero, and one may omit the first two entries of the
  list (1).}, hence they form a basis of a vector space. We now give 
a different set of $9$ differential operators, whose elements have a much 
simpler behaviour under $(\nu,x_2)\leftrightarrow(\mu,y)$:
\begin{align}\label{eq:basis2}
	&\text{(1)}
	&& g^{\mu\nu}\sum_{k} \square_{k}
	&&\sum_{k} \partial_k^{\mu}\partial_k^{\nu}
	&& \partial_{2}^{\mu} \partial_{y}^{\nu}
	&& \partial_{y}^{\mu} \partial_{2}^{\nu}
	&& g^{\mu\nu} \partial_{y}^{\alpha} \partial_{2\alpha}
\nonumber\\
	&\text{(2)}
	&& g^{\mu\nu} \square_{2}
	&& g^{\mu\nu} \square_{y}
	&& \partial_{2}^{\mu} \partial_{2}^{\nu}
	&& \partial_{y}^{\mu} \partial_{y}^{\nu} \; ,
\end{align}
where again $k \in \{11,...,1m\}$. By using
$$
\sum_k \partial_k = -\partial_2-\partial_y  \qquad 
	\text{and} \qquad \sum_{k\neq l}\del_k\del_l
= \Big(\sum_k \partial_k\Big)^2 - \sum_k \partial_k \partial_k \;,
$$
we can express all elements of the (old) basis \eqref{eq:basis1} as linear 
combination of the new terms \eqref{eq:basis2}; therefore, the latter are also
a basis of the same vector space.

Under $(\nu,x_2)\leftrightarrow(\mu,y)$, all differential operators in the 
group (1) (first line of \eqref{eq:basis2}) are individually invariant; hence, 
the pertinent contributions to $u^{\mu\nu}_n$ are admissible finite 
renormalizations. 

To treat the remaining four terms in group (2) (second line of 
\eqref{eq:basis2}), we proceed analogously to \eqref{eq:dd-2c}: we claim that
\be\label{eq:ddu}
\del_\nu^{x_2} \partial_{\mu}^y u^{\mu\nu}_n(L,\ldots,L,j^{\nu})
(x_{11}-y,\ldots,x_{1m}-y,x_2-y)
\word{is invariant under $x_2\leftrightarrow y$.}
\ee
To verify this, we insert \eqref{eq:case1} into \eqref{eq:ddu}:
obviously, the $\del_{x_2}\del_y t_{n+1}(\ldots,j,j)$-term and the 
$\del_{x_2}\del_y \bigl(\delta(y-x_2)\,t_n(\ldots,\phi^*\phi)\bigr)$-term
fulfil the claim individually. To show this for the remaining term, we
use the MWI to order $(n-1)$:
\begin{align*}
&-\sum_{l=1}^m\del_{\mu}^y\dl(y-x_{1l})\,\del_\nu^{x_2}
t_n\bigl(L(x_{11}),\dots,(\th^\mu L)(x_{1l}),\ldots,j^\nu(x_2)\bigr)=\\
&\qquad\sum_{k\neq l} \del_{\mu}^y\dl(y-x_{1l})\,\del_\nu^{x_2}\dl(x_2-x_{1k})\,
t_{n-1}\bigl(L(x_{11}),\dots,(\th^\mu L)(x_{1l}),\ldots,(\th^\nu L)(x_{1k}),
\ldots,L(x_{1m})\bigr),
\end{align*}
from which we see that also this term satisfies the claim \eqref{eq:ddu}.

We conclude that the contribution from the four terms in group (2) (second line of 
\eqref{eq:basis2}) satisfies
$$
0=\del_\nu^2 \partial_{\mu}^y\Bigl(C_1\, g^{\mu\nu} \square_{2}+C_2\,
g^{\mu\nu} \square_{y}+C_3\,\partial_{2}^{\mu} \partial_{2}^{\nu}+C_4\,
\partial_{y}^{\mu} \partial_{y}^{\nu}\Bigr)-(x_2\leftrightarrow y).
$$
Working out this condition, we obtain the relation
$$
C_1 = C_2 - C_3 + C_4.
$$
Thus eliminating $C_1$, we find that the remaining anomaly has the most 
general form 
$$
d_n^{\nu}(\dots)=\partial_{\mu}^y \Big(C_2\, g^{\mu\nu}(\square_{y} + \square_{2})+ 
	C_3\,(\partial_{2}^{\mu} \partial_{2}^{\nu} - g^{\mu\nu} \square_{2})
	+ C_4\, (\partial_{y}^{\mu} \partial_{y}^{\nu} +g^{\mu\nu} \square_{2})\,
	\Big) \, \delta(\dots)\; .
$$
At this point, we  exploit the freedom to change $u^{\mu\nu}$ without changing
  $\partial^y_\mu u^{\mu\nu}$. This allows us to replace $C_2(\dots) +
  C_3(\dots)+C_4(\dots)$ by
$$
(C_2+C_4)\, g^{\mu\nu}(\square_{y} + \square_{2})+ 
	C_3\,(\partial_{2}^{\mu} \partial_{2}^{\nu} - g^{\mu\nu}
        \square_{2}+\partial_{y}^{\mu} \partial_{y}^{\nu} - g^{\mu\nu}
        \square_{y}),
$$
      which still cancels the anomaly, and enjoys the required
      symmetry under $(y,\mu)\leftrightarrow(x_2,\nu)$. $\qed$

\section{Relation between the Master Ward identity and
its simplified version}\label{sec:MWI->WI}

Our aim is to generally relate the time-ordered products $T$ and $\wh T$
and to establish that the validity of the MWI for $T$ \eqref{eq:MWI} 
is equivalent to the validity of the WI for $\wh T$ \eqref{eq:WI-hat}
-- to all orders and including all loop diagrams.

We assume that $B_1,\ldots,B_n\in\sP$ are eigenvectors of
$\th$ \eqref{eq:theta},
\be\label{eq:theta-EW}
\th B_j=b_j\,B_j,\word{with eigenvalues} b_j\in\bZ,\quad\forall 1\leq j\leq n,
\ee
and that each of these field polynomials contains at most one 
derivated basic field, that is,
\be\label{eq:assume}
\frac{\del^2 B_j}{\del(\del^\mu\phi)\,\del(\del^\nu\phi)}=0,\quad
\frac{\del^2 B_j}{\del(\del^\mu\phi^*)\,\del(\del^\nu\phi)}=0,\quad
\frac{\del^2 B_j}{\del(\del^\mu\phi^*)\,\del(\del^\nu\phi^*)}=0,\quad
\forall 1\leq j\leq n.
\ee
Obviously these two assumptions are true for all elements of the 
set $\sP_0$, for which we have proved the validity of the MWI in 
Sect.~\ref{sec:MWI-proof}.

\subsection{Complete definition of the finite renormalization.}

In \cite{DKS93} it was investigated how the addition $ig^{\mu\nu}\dl$ to
$\del^\mu\del^\nu\Dl^F$ propagates to higher orders in the inductive 
Epstein--Glaser construction of the sequence $(\wh T_n)$. This was done
there only for tree-like diagrams; more precisely, diagrams consisting of two
components which are connected only by one internal $\phi$-line with two 
derivatives, and the consequences of the addition $ig^{\mu\nu}\dl$ to this line 
were studied. By virtue of the Main Theorem of Renormalization 
(\cite[Eq.~(3.6.25)]{D19} and Thm.~\ref{th:main-thm-renorm}), 
we are able to give a general definition of the higher
orders $\wh T\equiv (\wh T_{n})$ -- in particular, the inner
$\phi$-line with two derivatives may be part of a loop. 

In fact, we shall do more, by showing in Thm.~\ref{thm:WI-MWI} the
equivalence of a one-parameter family of Ward identities for a family
of time-ordered products $\wh T_c$, continuously
interpolating between $T=\wh T_{c=0}$ and $\wh T=\wh T_{c=1}$. The stronger
result for all $c\in\bR$ was suggested by the analogous result 
\cite[Sect.~3.3]{Tipp19} found at tree-level (where all
  renormalizations are fixed by the time-ordered 2-point functions of
  the derivative fields, and the absence of anomalies can be seen
  explicitly) in a different but presumably 
equivalent setup of scalar QED, using ``string-localized'' potentials.

To interpolate between the time-ordered products $T$ and $\wh T$, we multiply
the addition $ig^{\mu\nu}\delta(x-y)$ in the finite renormalization 
\eqref{eq:fin-ren} by a number $c\in\bR$, and denote the interpolating 
time-ordered product by $\wh T_c$. To formulate completely the
so-modified finite renormalization --
in particular its consequences for the higher orders in the inductive
Epstein--Glaser construction of $\wh T_c\equiv (\wh T_{c,n})$ -- we work
with a finite renormalization map $Z_c\equiv (Z_c^{(k)})$, which is an element 
of a version of the St\"uckelberg-Petermann renormalization group
(defined in Appendix \ref{sec:SP-RG-def}), and we will use 
the Main Theorem of Renormalization (given also in that Appendix in Theorem 
\ref{th:main-thm-renorm}).

To first order an element $Z_c$ of the St\"uckelberg-Petermann 
group is given by $Z_c^{(1)}\bigl(B(x)\bigr):=B(x)$ for all $B\in\sP$.
 
To second order, $\wh T_{c,2}$ differs from $T_2$ only by the finite 
renormalization $\del^\nu\del^\mu\Dl^F\to\del^\nu\del^\mu\Dl^F+c\,ig^{\mu\nu}\dl$
in the connected tree-diagram part; hence we define
\begin{align}
Z_c^{(2)}\bigl(B_1(x_1),B_2(x_2)\bigr):=&\,c\,i\bigl(
\wh T_2\bigl(B_1(x_1),B_2(x_2)\bigr)-T_2\bigl(B_1(x_1),B_2(x_2)\bigr)\bigr)
\label{eq:Z2-general}\\
\equiv &\,c\,\zeta(B_1,B_2)(x_1)\,\dl(x_1-x_2)\label{eq:Z2}\\
\word{where}
\zeta(B_1,B_2):=&\,
\frac{\del B_1}{\del(\del^\mu\phi^*)}\,\frac{\del B_2}{\del(\del_\mu\phi)}
+\frac{\del B_1}{\del(\del^\mu\phi)}\,\frac{\del B_2}{\del(\del_\mu\phi^*)}.
\label{eq:zeta}
\end{align}
For later purpose we note that
\be\label{eq:th-chi}
\th\,\zeta(B_1,B_2)=(b_1+b_2)\,\,\zeta(B_1,B_2).
\ee
For $n\geq 3$ the difference between $\wh T_{c,n}$ and $T_n$ is only the
one coming from the propagation of $Z_c^{(2)}$ to higher orders; hence the
higher orders of $Z_c$ vanish, that is,
\be\label{eq:Zk=0}
Z_c^{(k)}\bigl(B_1(x_1)\oxyox B_k(x_k)\bigr)=0\quad\forall k\geq 3.
\ee
We point out: $Z_c^{(2)}$ does not fulfil the AWI and the property 
``Field Equation'', because $\wh T_{2,c}$ violates these relations.
Comparing with the definition of the St\"uckelberg-Petermann 
group in the mentioned references, the $Z_c$ defined above is an element 
of a \emph{modified version} of that group; this is explained in detail
in parts \ref{sec:SP-RG-def}--\ref{sec:Z2inR} of the Appendix.

We generally define $\wh T_c\equiv (\wh T_{c,n})$ in terms of $T$ and $Z_c$ 
by using \cite[Eq.~(3.6.25)]{D19}:
\begin{align}\label{eq:MainThm-n}
i^n\,\wh T_{c,n}\bigl(\ox_{j=1}^n B_j(x_j)\bigr):=& i^n\, 
T_n\bigl(\ox_{j=1}^n B_j(x_j)\bigr)\nonumber\\
&+\sum_{\substack{P\in\mathrm{Part}_2(\{1,\dots,n\})\\ n/2\leq|P|<n}}
i^{|P|}\,T_{|P|}\left(\bigox_{I\in P} Z_c^{(|I|)}\bigl(\ox_{j\in I}B_j(x_j)\bigr)\right),
\end{align}
where $P\in\mathrm{Part}_2(\{1,\dots,n\})$ is a partition of $\{1,\ldots,n\}$
into $|P|$ disjoint subsets $I$, each of these subsets has $|I|=1$ or $|I|=2$
elements. (The latter is the reason for the subscript ``$2$'' in 
$\mathrm{Part}_2$.) The term $|P|=n$ is explicitly written out. For $n=2$ 
the formula \eqref{eq:MainThm-n} reduces to the general definition of 
$Z_c^{(2)}$ given in \eqref{eq:Z2-general}.

Part (b) of the Main Theorem of Renormalization 
states that the so-defined $\wh T_c$ is also a time-ordered product, that is, it satisfies the basic axioms and the 
renormalization conditions (v)-(viii) given in Appendix \ref{sec:axioms-T};
however, it may violate the AWI, the
FE and any Ward identities. The proof of this statement
is given in Appendix \ref{sec:hat-T}.

\begin{remk}
Renormalizing the interaction $e(jA)(g)$ by the given SP renormalization map 
$Z_c$ (according to \eqref{eq:renorm-int}), we indeed obtain 
$e(jA)(g)+c\,e^2(AA\phi^*\phi)(g^2)$ (where $g^2(x):=(g(x))^2$). 
In detail we get 
\begin{align}
\sZ_c\bigl((g,ejA)\bigr)= &\, e(jA)(g)+\frac{e^2}{2!}\int dx_1dx_2\,\,
g(x_1)g(x_2)\,Z^{(2)}_c\bigl((jA)(x_1),(jA)(x_2)\bigr)\nonumber\\
= &\, e(jA)(g)+c\,e^2(AA\phi^*\phi)(g^2).\label{eq:Z(Ww)}
\end{align}
As explained at the end of Appendix \ref{sec:SP-RG-def},
the above definition of $\wh T_c$ \eqref{eq:MainThm-n} can be
written in terms of the $S$-matrix \eqref{eq:S-matrix} by means of the formula
\eqref{eq:hatS=SZ}: $\wh\nS_c:=\nS\circ \sZ_c$ by abuse of notation. 
For the interaction $\tilde L:=e\,j^\mu A_\mu$ \eqref{eq:tilde-L} this yields
\be
\wh\nS_c\bigl((g,ejA)\bigr):=\nS\bigl((g,ejA),(g^2,ce^2AA\phi^*\phi)\bigr)
\quad\forall g\in\sD(\bM),
\ee
by using \eqref{eq:Z(Ww)}. For $c=1$, this is precisely the relation between
$\wh T$ and $T$ we want to hold -- see \eqref{eq:L-ren}.

The physically relevant $S$-matrix for scalar QED, that is,
$\nS\bigl((g,ejA),(g^2,e^2AA\phi^*\phi)\bigr)$, can be expressed in terms of 
$\wh\nS_c$ by
\be\label{eq:scal-QED-S}
\nS\bigl((g,ejA),(g^2,e^2AA\phi^*\phi)\bigr)=
\wh\nS_c\bigl((g,ejA),(g^2,(1-c)e^2AA\phi^*\phi)\bigr), 
\quad\forall c\in\bR,\,g\in\sD(\bM).
\ee
This relation is obtained by applying $\wh\nS_c:=\nS\circ \sZ_c$ to 
$\bigl((g,ejA),(g^2,(1-c)e^2AA\phi^*\phi)\bigr)$, to wit, by using the explicit formulas for
$Z_c$ \eqref{eq:Z2}--\eqref{eq:Zk=0} we get
\begin{align*}
\sZ_c&\bigl((g,ejA),(g^2,(1-c)e^2AA\phi^*\phi)\bigr)\\
=&e(jA)(g)+(1-c)e^2(AA\phi^*\phi)(g^2)+\frac{e^2}{2!}\int dx_1dx_2\,\,
g(x_1)g(x_2)\,Z^{(2)}_c\bigl((jA)(x_1),(jA)(x_2)\bigr)\\
=&e(jA)(g)+e^2(AA\phi^*\phi)(g^2).
\end{align*}
\end{remk}

\begin{remk}[Interacting electromagnetic current in terms of $\wh T$]
Working with the time-ordered product $\wh T\equiv \wh T_{c=1}$, 
Bogoliubov's definition \eqref{eq:int-field} of the interacting electromagnetic current reads
\be\label{eq:curr-hat}
\wh{j^\mu_{(g,\tilde L),0}}(\al):=\wh{\nS}\bigl((g,\tilde L)\bigr)_0^{\star -1}
\star \frac{d}{i\,d\la}\Big\vert_{\la=0}
\wh{\nS}\bigl((g,\tilde L),(\al,\la j^\mu)\bigr)_0,\quad g,\al\in\sD(\bM,\bR),
\ee
where $\tilde L:=ejA$ \eqref{eq:tilde-L}. We are going to show that the 
definition $\wh\nS:=\nS\circ \sZ$ \eqref{eq:hatS=SZ} (where $\sZ\equiv \sZ_{c=1}$) implies that
\be\label{eq:j=J}
\wh{j^\mu_{(g,\tilde L),0}}(x)=J^\mu_S(x)_0,
\ee
where $J^\mu$ is given in \eqref{eq:curr-cons} and
$S:=e(j^\mu A_\mu)(g)+e^2(A^2\phi^*\phi)(g^2)$ \eqref{eq:interaction}. 
To do this we insert $\wh\nS:=\nS\circ \sZ$ into 
\eqref{eq:curr-hat} and use that $\sZ\bigl((g,ejA)\bigr)=S$ \eqref{eq:Z(Ww)}.
This yields
\be\label{eq:j-hat}
\wh{j^\mu_{(g,\tilde L),0}}(\al)=\nS(S)_0^{\star -1}\star 
T\Bigl(e_\ox^{iS}\ox \frac{d}{d\la}\Big\vert_{\la=0}
\sZ\bigl((g,\tilde L),(\al,\la j^\mu)\bigr)\Bigr)
\ee
Using \eqref{eq:renorm-int} and the explicit formulas for $Z$ we obtain
\begin{align*}
\frac{d}{d\la}\Big\vert_{\la=0}\sZ\bigl((g,\tilde L),(\al,\la j^\mu)\bigr)=&
j^\mu(\al)+e\int dx_1dx_2\,\,g(x_1)\al(x_2)\,Z^{(2)}\bigl((jA)(x_1),j^\mu(x_2)\bigr)
\nonumber\\
=&j^\mu(\al)+2e\,(A^\mu\phi^*\phi)(g\al)=J^\mu(\al).
\end{align*}
Inserting this result into \eqref{eq:j-hat} and comparing with the definition
of $J^\mu_S(\al)_0$ \eqref{eq:int-field}, we get the assertion \eqref{eq:j=J}.

The equality \eqref{eq:j=J} can be understood in terms of Feynman
diagrams: there are diagrams contributing to the second factor on the
r.h.s.~of \eqref{eq:curr-hat}, i.e.
$\wh T\bigl((\ox_k\tilde L(y_k))\ox j^\mu(x)\bigr)_0$, in which the field 
vertex $x$ is connected to an interaction vertex $y_k$ by an internal 
$\phi$-line with two derivatives and this line is not part of 
any loop. The 
addition $ig^{\mu\nu}\dl$ to this line in these diagrams generates the
additional term $2eg(x)(A^\mu\phi^*\phi)_S(x)_0$ of $J^\mu_S(x)_0$.

From the identity \eqref{eq:j=J} and $\del^x_\mu J^\mu_S(x)_0=0$
\eqref{eq:curr-cons} (that is secured by the assumed validity of the MWI
    for the T-products $T$), we see that $\wh{j^\mu_{(g,\tilde L),0}}$ is
conserved:
$$
\del_\mu^x\wh{j^\mu_{(g,\tilde L),0}}(x)=0.
$$ 
Alternatively, this result can directly be obtained, i.e., without using 
$J^\mu_S$; namely, from the WI for $\wh T$ \eqref{eq:WI-hat},
by proceeding analogously to the derivation of 
$\del_\mu^x J^\mu_S(x)_0=0$ from the MWI \eqref{eq:MWI-F}. 
Hence, with regard to the interacting electromagnetic 
current the WI \eqref{eq:WI-hat} and the MWI \eqref{eq:MWI} 
(or \eqref{eq:MWI-F}) contain the same information. This result can 
strongly be generalized -- this is the topic of the next subsection.
\end{remk}

\subsection{The MWI for $T$ \eqref{eq:MWI} and the WI for $\wh T$
  \eqref{eq:WI-hat} are equivalent}
\label{sec:WI-MWI}
We are now coming to the second main result of this paper.

\begin{thm}\label{thm:WI-MWI}
Given a time-ordered product $T$ and  $c\in\bR$, let a time-ordered product
$\wh T_c$ be defined in terms of $T$ and 
$Z_c$ by \eqref{eq:Z2-general}--\eqref{eq:zeta} and \eqref{eq:MainThm-n}.
Then, for all $B_j\in\sP$ satisfying the assumptions \eqref{eq:theta-EW}
and \eqref{eq:assume}, the validity of the MWI 
\eqref{eq:MWI} for $T$ is equivalent to the validity of the
following $c$-dependent WI for $\wh T_c$ -- to all orders $n\in\bN$:
\begin{align}\label{eq:WI-c}
\del_y^\mu\,\wh T_{c,n+1}\bigl(B_1(x_1)&\ox\cdots\ox  B_n(x_n)\ox
j_\mu(y)\bigr)_0=\\
&\sum_{l=1}^n\dl(y-x_l)\,\wh T_{c,n}\bigl( B_1(x_1)\ox \cdots
\ox(\th B_l)(x_l)\ox\cdots\ox  B_n(x_n)\bigr)_0\nonumber\\
&+(c-1)\,\del^\mu_y\Bigl(\sum_{l=1}^n\dl(y-x_l)\,
\wh T_{c,n}\bigl( B_1(x_1)\ox \cdots
\ox(\th_\mu B_l)(x_l)\ox\cdots\ox  B_n(x_n)\bigr)_0\Bigr),\nonumber
\end{align}
\end{thm}

For $c=1$ the assertion \eqref{eq:WI-c} agrees with the WI \eqref{eq:WI-hat} 
and for $c=0$ with the MWI \eqref{eq:MWI}. In particular we obtain
\begin{align}\label{eq:WI-c-examp}
&\del^y_\mu\,\wh T_{c,2}\bigl(\del^\nu\phi(x)\ox j^\mu(y)\bigr)_0=
\dl(y-x)\,\del^\nu\phi(x)_0+(c-1)\,(\del^\nu\dl)(y-x)\,\phi(x)_0,\nonumber\\
&\del^y_\mu\,\wh T_{c,2}\bigl(\del^\nu\phi^*(x)\ox j^\mu(y)\bigr)_0=
-\dl(y-x)\,\del^\nu\phi^*(x)_0-(c-1)\,(\del^\nu\dl)(y-x)\,\phi^*(x)_0,
\nonumber\\
&\del^y_\mu\,\wh T_{c,2}\bigl(j^\nu(x)\ox j^\mu(y)\bigr)_0=
(1-c)\,2i\,(\phi^*\phi)(x)_0\,\del^\nu\delta(y-x),
\end{align}
which contains the relations \eqref{eq:WI-hat-2} for $\wh T=\wh T_{c=1}$ and
\eqref{eq:MWI-2} for $T=\wh T_{c=0}$.

\begin{proof} \emph{MWI \eqref{eq:MWI} for $T$ $\Longrightarrow$ 
WI \eqref{eq:WI-c} for $\wh T_c$:}
From \eqref{eq:Z2}--\eqref{eq:zeta} we obtain 
\be\label{eq:Z(Bj)}
Z_c^{(2)}\bigl(B(x),j^\mu(y)\bigr)=ic\,(\th^\mu B)(x)\,\dl(y-x),
\ee
with $\th^\mu$ defined in \eqref{eq:th-mu}.

Using \eqref{eq:MainThm-n} to express $\wh T_c$ in terms of $T$ on the l.h.s.\ of \eqref{eq:WI-c}, 
we get two types of terms: in the first type $j^\mu$ does not appear in the 
argument of any $Z_c^{(2)}$, in the second type it does and, hence, we may use
\eqref{eq:Z(Bj)}:
\begin{align} 
&i^{n+1}\text{[l.h.s.\ of WI]}=
\sum_{\substack{P\in\mathrm{Part}_2(\{1,\dots,n\})\\ n/2\leq|P|\leq n}}
i^{|P|+1}\,\del_\mu^y T_{|P|+1}\left(\bigox_{I\in P} Z_c^{(|I|)}\bigl(\ox_{j\in I}B_j(x_j)
\bigr),j^\mu(y)\right)_0\label{eq:L1}\\
&+ic\sum_{l=1}^n(\del_\mu\dl)(y-x_l)
\sum_{\substack{Q\in\mathrm{Part}_2(\{1,\dots,\hat l,\ldots,n\})\\ (n-1)/2\leq|Q|\leq n-1}}
i^{|Q|+1}\, T_{|Q|+1}\left((\th^\mu B_l)(x_l)\ox
\bigl[\bigox_{I\in Q} Z_c^{(|I|)}\bigl(\ox_{j\in I}B_j(x_j)
\bigr)\bigr]\right)_0,\label{eq:L2}
\end{align}
where $\hat l$ means that $l$ is omitted in the pertinent set.
Now we insert the MWI \eqref{eq:MWI} into \eqref{eq:L1}: for the
$\th$-terms (displayed in \eqref{eq:L1A}) we use \eqref{eq:theta-EW} and
\eqref{eq:Z2}--\eqref{eq:th-chi}, the latter imply
\begin{align*}
\dl(y-x_k)\,&T\bigl(\cdots\ox\th Z_c^{(2)}\bigl(B_k(x_k),B_j(x_j)\bigr)
\ox\cdots\bigr)\\
&=\dl(y-x_k,y-x_j)\,(b_j+b_k)\,
T\bigl(\cdots\ox c\,\zeta(B_k,B_j)(x_k)\ox\cdots\bigr)\\
&=\bigl[\dl(y-x_k)\,b_k+\dl(y-x_j)\,b_j\bigr]\cdot
T\bigl(\cdots\ox Z_c^{(2)}\bigl(B_k(x_k),B_j(x_j)\bigr)\ox\cdots\bigr).
\end{align*}
For the $\th^\mu$-terms (displayed in \eqref{eq:L1B}) we take into account 
that $\th^\mu\,\zeta(B_1,B_2)=0$, which follows from \eqref{eq:assume},
and we reorder the summations. So we obtain:
\begin{align}
&\text{\eqref{eq:L1}}
=i\left[\sum_{\substack{P\in\mathrm{Part}_2(\{1,\dots,n\})\\ n/2\leq|P|\leq n}}
i^{|P|}\,T_{|P|}\left(\bigox_{I\in P} Z_c^{(|I|)}\bigl(\ox_{j\in I}B_j(x_j)\bigr)
\right)_0\right]\cdot\left[\sum_{l=1}^n\dl(y-x_l)\,b_l\right]\label{eq:L1A}\\
&-i\sum_{l=1}^n(\del_\mu\dl)(y-x_l)
\sum_{\substack{Q\in\mathrm{Part}_2(\{1,\dots,\hat l,\ldots,n\})\\ (n-1)/2\leq|Q|\leq n-1}}
i^{|Q|+1}\, T_{|Q|+1}\left((\th^\mu B_l)(x_l)\ox
\bigl[\bigox_{I\in Q} Z_c^{(|I|)}\bigl(\ox_{j\in I}B_j(x_j)\bigr)\bigr]\right)_0.
\label{eq:L1B}
\end{align}

Finally, we reexpress $T$ in terms of $\wh T_c$. 
Thanks to \eqref{eq:assume} and \eqref{eq:Z2}--\eqref{eq:zeta}, it holds that
$$
Z_c^{(2)}\bigl((\th^\mu B_l)(x_l)\ox B_j(x_j)\bigr)=0.
$$
Hence, we obtain
\begin{align*}
&\sum_{\substack{Q\in\mathrm{Part}_2(\{1,\dots,\hat l,\ldots,n\})\\ (n-1)/2\leq|Q|\leq n-1}}
i^{|Q|+1}\, T_{|Q|+1}\left((\th^\mu B_l)(x_l)\ox
\bigl[\bigox_{I\in Q} Z_c^{(|I|)}\bigl(\ox_{j\in I}B_j(x_j)\bigr)\bigr]\right)_0\\
&\qquad\qquad\qquad =i^n\,\wh T_{c,n}\bigl(B_1(x_1)\ox \cdots\ox(\th^\mu B_l)(x_l)
\ox\cdots\ox  B_n(x_n)\bigr)_0.
\end{align*}
So we see that the sum of the terms \eqref{eq:L2} and \eqref{eq:L1B} 
is equal to $i^{n+1}\cdot$[$(c-1)$-term on the r.h.s.\ of the assertion
\eqref{eq:WI-c}]. And, the expression \eqref{eq:L1A} is equal to
\begin{align*}
&i^{n+1}\,\wh T_{c,n}\bigl(\ox_{j=1}^n B_j(x_j)\bigr)_0
\cdot\left[\sum_{l=1}^n\dl(y-x_l)\,b_l\right]\\
&\qquad =i^{n+1}\sum_{l=1}^n\dl(y-x_l)\,\wh T_{c,n}\bigl( B_1(x_1)\ox \cdots
\ox(\th B_l)(x_l)\ox\cdots\ox  B_n(x_n)\bigr)_0,
\end{align*}
by using \eqref{eq:theta-EW}.

\emph{WI \eqref{eq:WI-c} for $\wh T_c$ $\Longrightarrow$ 
  MWI \eqref{eq:MWI} for $T$:} First we show that, 
for $B_j$'s satisfying \eqref{eq:assume}, the ``inverse'' of $Z_c\in\sR$ 
(see Def.~\ref{df:SP-RG} for the definition of $\sR$) is 
$Y_c\in\sR$ given by  $Y_c^{(1)}\bigl(B(x)\bigr):=B(x)$ and 
\begin{align*}
&Y_c^{(2)}\bigl(B_1(x_1),B_2(x_2)\bigr):=-c\,\zeta(B_1,B_2)(x_1)\,\dl(x_1-x_2),\\
&Y_c^{(k)}\bigl(B_1(x_1),\ldots,B_k(x_k)\bigr):=0\quad\forall k\geq 3.
\end{align*}
Since $\sZ_c\bigl((g_j,B_j)\bigr)=\sum_jB_j(g_j)+\tfrac{c}2\sum_{j_1,j_2}\zeta(B_{j_1},B_{j_2})(g_{j_1}g_{j_2})$, 
we have to verify the relation
$\sY_c\bigl((g_j,B_j),(g_{j_1}g_{j_2},\tfrac{c}2\,\zeta(B_{j_1},B_{j_2}))\bigr)=\sum_jB_j(g_j)$, 
where $\sY_c$ denotes the renormalization 
of the interaction given by $Y_c\in\sR$ according to \eqref{eq:renorm-int}.
Taking into account that $\zeta\bigl(B_{j_1},\zeta(B_{j_2},B_{j_3})\bigr)=0=
\zeta\bigl(\zeta(B_{j_1},B_{j_2}),\zeta(B_{j_3},B_{j_4})\bigr)$ 
because $\frac{\del\zeta(B_{j_1},B_{j_2})}{\del(\del^\mu\phi)}=0=
\frac{\del\zeta(B_{j_1},B_{j_2})}{\del(\del^\mu\phi^*)}$,  we indeed obtain
\begin{align*}
Y_c\bigl((g_j,B_j)&,(g_{j_1}g_{j_2},\tfrac{c}2\,\zeta(B_{j_1},B_{j_2}))\bigr)=\sum_jB_j(g_j)+
\frac{c}2\sum_{j_1.j_2}\zeta(B_{j_1},B_{j_2})(g_{j_1}g_{j_2})\\
&+\frac{1}2\,\int dx_1dx_2\,\,g_{j_1}(x_1)g_{j_2}(x_2)\,
Y_c^{(2)}\bigl(B_{j_1}(x_1),B_{j_2}(x_2)\bigr)=\sum_jB_j(g_j).
\end{align*}
Therefore, $T_n$ can be expressed in terms of 
$(\wh T_{c,k})_{1\leq k\leq n}$ and $Y_c$ by 
the formula \eqref{eq:MainThm-n}: $T$ and $\wh T_c$ are mutually exchanged 
and $Z_c$ is replaced by $Y_c$.

With this, the assertion (i.e.~the MWI \eqref{eq:MWI} for $T$) can be 
verified by essentially the same computation as in the 
above proof of the reversed
statement: to compute $\del_\mu^y T_{n+1}\bigl(\cdots\ox j^\mu(y)\bigr)$ we
first express $T_{n+1}$ in terms of $\wh T_c$, then we use the WI 
\eqref{eq:WI-c} for $\wh T_c$ and finally we reexpress $\wh T_c$ in 
terms of $T$.
\end{proof}

\appendix

\section{St\"uckelberg--Petermann renormalization group 
without Action Ward Identity}
\label{app:SP-RG}

\subsection{Axioms for the time-ordered product}
\label{sec:axioms-T}

Both time-ordered products $T$ and $\wh T$, used in the main text, 
satisfy the following definition:

\begin{defn}\label{df:T}
A time-ordered product $T$ is a sequence of maps (61)$\to$ Sect.~1.1 
\be\label{eq:hatT}
T\equiv (T_n)_{n=1}^\infty\:\begin{cases}
 \sP^{\ox n} \longto \sD'(\bM^n,\sF)\\
B_1\oxyox B_n\longmapsto  T_n\bigl(B_1(x_1)\oxyox B_n(x_n)\bigr),\end{cases}
\ee
fulfilling certain axioms -- the basic axioms and the renormalization 
conditions. The former read:
\begin{enumerate}
\item[(i)]  \textbf{Linearity}:
$T_n$ is \emph{linear} (that is, multilinear in $(B_1,\ldots,B_n)$);
\item[(ii)]  \textbf{Initial Condition}:
$T_1\bigl(B(x)\bigr) = B(x)$ for any $B \in \sP\,$;
\item[(iii)] \textbf{Symmetry}: For all permutations 
$\pi$ of $(1,\ldots,n)$ it holds that
$$
T_n\bigl(B_{\pi 1}(x_{\pi 1})\oxyox B_{\pi n}(x_{\pi n})\bigr)=
T_n\bigl(B_1(x_1)\oxyox B_n(x_n)\bigr).
$$
\item[(iv)]  \textbf{Causality}.
For all $B_1,\ldots,B_n\in\sP$, $T_n$ fulfills the causal factorization:
$$ 
T_n\bigl(B_1(x_1),\dots,B_n(x_n)\bigr)
= T_k\bigl(B_1(x_1),\dots,B_k(x_k)\bigr) \star
T_{n-k}\bigl(B_{k+1}(x_{k+1}),\dots, B_n(x_n)\bigr)
$$
whenever
$\{x_1,\dots,x_k\} \cap \bigl(\{x_{k+1},\dots,x_n\}+\ovl V_-\bigr)
= \emptyset\,$ where $\ovl V_-$ is the closed backward lightcone.
\end{enumerate}
We work with the following renormalization conditions:
\begin{enumerate}
\item[(v)] \textbf{Field Independence}: 
$$
\fd{T_n\bigl(B_1(x_1)\oxyox B_n(x_n)\bigr)}{z}=\sum_{j=1}^n 
T_n\bigl(B_1(x_1)\oxyox \fd{B_j(x_j)}{z}\oxyox B_n(x_n)\bigr)
$$
and similarly for $\frac{\dl}{\dl\phi^*(z)}$ and $\frac{\dl}{\dl A^\mu(z)}$.
This axiom is equivalent to the requirement that
$T_n$ satisfies the \emph{causal Wick expansion}, which is a \emph{unique} prescription for the expansion of the time-ordered product in terms of 
Wick products. To wit,
for \textit{monomials} $B_1,\dots,B_n\in\sP$ it holds that
\begin{equation}
T_{n}\bigl( B_1(x_1),\dots, B_n(x_n) \bigr)
= \sum_{\unl B_l \subset B_l}\! \om_0\bigl( 
T_{n}\bigl( \unl B_1(x_1),\dots, \unl B_n(x_n) \bigr) \bigr)\,
\ovl B_1(x_1) \cdots \ovl B_n(x_n),
\label{eq:causal-Wick-expan}
\end{equation} 
where the \textit{submonomials} $\unl B$ of a given monomial $B\in\sP$ of
order $n$ and their
\textit{complementary submonomials} $\ovl B$ are defined by
\begin{align}\label{eq:submonomials}
\unl B
   &:= \frac{\del^k B}{\del\vf_{a_1}\cdots\del \vf_{a_k}}
\neq 0,\\
\ovl B 
&:= C_{a_1\dots a_k} \,\vf_{a_1} \cdots \vf_{a_k}
\quad \text{(no sum over $a_1,\dots,a_k$)},\nonumber
\end{align}
where each $C_{a_1\dots a_k}$ is a certain combinatorial factor. The range of the sum $\sum_{\unl B \subset B}$ are all allowable
$k\leq n$ and indices $a_1,\dots,a_k$ of the fields
$\vf_a=A^\mu,\phi,\phi^*,\del^\mu\phi$ and $\del^\nu\phi^*$ generating
$\sP$. (For $k = 0$ we have $\unl B = B$ and
$\ovl B = 1$.)
\item[(vi)] \textbf{$*$-Structure}: To formulate this axiom, we 
introduce the $S$-matrix to the interaction
$$
\sum_{j=1}^JB_j(g_j)\equiv\sum_{j=1}^J\int dx\,\,B_j(x)\,g_j(x),
\quad B_j\in\sP,\,\,g_j\in\sD(\bM);
$$
it is the generating functional of the time-ordered products,
understood as a formal series in the $g_j$'s:
\begin{align}\label{eq:S-matrix}
&\nS\bigl((g_j,B_j)_{j=1}^J\bigr):= \\
&1 + \sum_{n=1}^\infty \frac{i^n}{n!}
\int dx_1\cdots dx_n\,\,\sum_{j_1,\ldots,j_n=1}^J
g_{j_1}(x_1)\cdots g_{j_n}(x_n)\, T_n\bigl(B_{j_1}(x_1),\dots,B_{j_n}(x_n)\bigr).
\nonumber
\end{align}
The axiom $*$-Structure reads
$$ 
\nS\bigl((g_j,B_j)_{j=1}^J\bigr)^*=
\nS\bigl((\ovl{g_j},B_j^*)_{j=1}^J\bigr)^{\star -1}
$$
where $G^{\star -1}$ is the inverse w.r.t.\ the star product of $G\in\sF$. For
a real interaction (i.e., $\ovl{g_j}=g_j,\,B_j^*=B_j$ for all $j$)
this axiom asserts the unitarity of the $S$-matrix
    w.r.t.\ the star product (and as a formal series).
\item[(vii)] \textbf{Poincar\'e Covariance}:
$$
\bt_{\La,a} T_n\bigl(B_1(x_1)\oxyox B_n(x_n)\bigr) = 
T_n\bigl(\bt_{\La,a} B_1(x_1)\oxyox \bt_{\La,a} B_n(x_n)\bigr)\quad
\forall (\La,a)\in\sP_+^\up,
$$
where $(\La,a)\mapsto\bt_{\La,a}$ is the natural representation of $\sP_+^\up$ 
on $\sF$ (see \cite[Chap.~3.1.4]{D19}). An immediate consequence of 
translation covariance is that the $\bC$-valued distributions 
\be\label{eq:tn}
t_n(B_1,\ldots,B_n)(x_1-x_n,\ldots):=\om_0\Bigl(
T_n\bigl(B_1(x_1)\oxyox B_n(x_n)\bigr)\Bigr)\in\sD'(\bM^{n-1},\bC)
\ee
depend only on the relative coordinates.

\item[(viii)]  \textbf{Scaling Degree}: With $\sd t$ denoting the scaling 
degree of the distribution $t$ w.r.t.\ the origin (see, 
e.g.,~\cite[Def.~3.2.5]{D19}) the VEVs \eqref{eq:tn} are required to fulfil
$$
\sd t_n(B_1,\ldots,B_n)(x_1-x_n,\ldots)\leq\sum_{j=1}^n\dim B_j
$$
for all $B_1,\ldots,B_n\in\sP_\homog$, where $\dim B$ is the mass dimension
of $B$ and $\sP_\homog$ is the subset of $\sP$ of all field polynomials being 
homogeneous in the mass dimension (see \cite[Chap.~3.1.5]{D19}).
\end{enumerate}
\end{defn}
 
The time-ordered product $T$, underlying Sects.~\ref{sec:MWI}, 
and \ref{sec:MWI-proof}, fulfils additionally the 
following two renormalization conditions.
\begin{enumerate}
\item[AWI] \textbf{Action Ward Identity}:
$$
\del_{x_j} T_{n}\bigl(B_1(x_1)\oxyox B_j(x_j)\ox\cdots\bigr)
=T_{n}\bigl(B_1(x_1)\oxyox \del_{x_j}B_j(x_j)\ox\cdots\bigr)
\quad\forall 1\leq j\leq n,
$$
which implies that $T_n$ can be interpreted as a map 
$T_n:\sF_\loc^{\ox n}\to\sF$ (or $T_n:\sF_\loc^{\oxsym n}\to\sF$ due to the axiom Symmetry,
with $\oxsym$ denoting the symmetrized tensor product); 
for details see \cite[Chap.~3.1.1]{D19}. For
$F_k=\sum_{j_k}\int dx\,\,g_{j_k}(x)\,B_{j_k}(x)\in\sF_\loc,\,\,g_{j_k}\in\sD(\bM),\,B_{j_k}\in\sP$; the two kinds of maps 
$T_n$ are related by
$$ 
T_n(F_1\oxyox F_n)=\sum_{j_1,\dots,j_n}\int dx_1\cdots dx_n\,\,g_{j_1}(x_1)\cdots g_{j_n}(x_n)\,
T_n\bigl(B_{j_1}(x_1)\oxyox B_{j_n}(x_n)\bigr)\ .
$$ 
\item[FE] \textbf{Field Equation}:
\begin{align*}
T_{n+1}\bigl(\del^a\phi(x)\ox B_1(x_1)\oxyox B_n(x_n)\bigr)=&
\del^a\phi(x)\,\,T_{n}\bigl(B_1(x_1)\oxyox B_n(x_n)\bigr)\\
+\int dy\,\,\del^a\Dl^F(x-y) & \,\frac{\dl}{\dl\phi^*(y)}
T_{n}\bigl(B_1(x_1)\oxyox B_n(x_n)\bigr)
\end{align*}
and analogously for $\phi$ replaced by $\phi^*$ or $A^\mu$.
\end{enumerate}

In the inductive step of the Epstein-Glaser construction of the sequence $(T_n)$, the basic axioms determine 
$T_n\bigl(B_1(x_1)\oxyox B_n(x_n)\bigr)$ uniquely on $\sD(\bM^n\less\Dl_n)$ in terms of the $T_k$ of lower orders $1\leq k<n$
(for all $B_1,\ldots,B_n\in\sP$), where 
\be\label{eq:diagonal}
\Dl_n:=\set{(x_1,\ldots,x_n)\in\bM^n\,\big\vert\,x_1=x_2=\ldots =x_n}
\ee 
is the thin diagonal. The extension of $T_n\bigl(B_1(x_1)\oxyox B_n(x_n)\bigr)$ to $\sD(\bM^n)$ is in general nonunique, 
it is named `renormalization', because it corresponds to what is `renormalization' in conventional approaches. The only
purpose of the further axioms (v)-(viii) (and possibly AWI and FE), is to restrict this extension; therefore, they are called 
`renormalization conditions'. They also give some indications how to do the extension:
thanks to the causal Wick expansion and translation covariance, the extension is done in terms of the numerical distributions $t_n$
\eqref{eq:tn}; hence the problem of renormalization is reduced to the extension of $\bC$-valued distributions to \emph{one point},
to wit, the extension from $\sD'(\bM^{n-1}\less\{0\},\bC)$ to $\sD'(\bM^{n-1},\bC)$. 
Two extensions $t_{n,1}$ and $t_{n,2}$  of
$t_n\in\sD'(\bM^{n-1}\less\{0\},\bC)$ differ by a finite linear combination of
derivatives of the $\delta$-distribution, the order of the derivatives being bounded by the renormalization condition Scaling Degree:
\be\label{eq:nonunique}
t_{n,2}(x)-t_{n,1}(x)=\sum_{|a|=0}^\om C_a\del_x^a\dl(x),\word{where}\om\equiv\om(t_n):=\sd(t_n)-4(n-1)
\ee
is called the `singular order' of $t_n$. The coefficients $C_a\in\bC$ are restricted by the further renormalization conditions.

\subsection{St\"uckelberg--Petermann renormalization group $\sR$ and 
Main Theorem of Renormalization}\label{sec:SP-RG-def}

In this paper, we work with that version of the 
St\"uckelberg--Petermann renormalization group (SP-RG) that describes finite 
renormalizations of time-ordered products satisfying the renormalization 
conditions (v)-(viii) given in the preceding Sect., however, they may violate 
the AWI and the FE. 
In the absence of the AWI, the arguments of the elements of the SP-RG
cannot be written as local functionals, as it is done in \cite{DF04,BDF09}
and \cite[Chap.~3.6]{D19}.

\begin{defn} 
\label{df:SP-RG}
The \emph{St\"uckelberg--Petermann renormalization group} 
is the set $\sR$ of all sequences of maps%
\footnote{Mind the difference: $Z^{(n)}$ takes values in the
$\sF_\loc$-valued distributions -- in contrast to $T_n$.}   
\be\label{eq:Z}
Z\equiv (Z^{(n)})_{n=1}^\infty\:\begin{cases}
 \sP^{\ox n} \longto \sD'(\bM^n,\sF_\loc)\\
B_1\oxyox B_n\longmapsto Z^{(n)}\bigl(B_1(x_1)\oxyox B_n(x_n)\bigr)\end{cases}
\ee
being \emph{linear} (that is, multilinear in $(B_1,\ldots,B_n)$) and 
\emph{symmetric} in the sense that
\be\label{eq:Z-sym}
Z^{(n)}\bigl(B_{\pi 1}(x_{\pi 1})\oxyox B_{\pi n}(x_{\pi n})\bigr)=
Z^{(n)}\bigl(B_1(x_1)\oxyox B_n(x_n)\bigr)
\ee
for all permutations $\pi$ of $(1,\ldots,n)$. In addition, the maps
$Z^{(n)}$ are required to satisfy the following properties for all
$B,B_1,\ldots,B_n\in\sP$ and for all $n\geq 1$:
\begin{itemize}
\item[(1)] \emph{Lowest order:} 
$Z^{(1)}\bigl(B(x)\bigr) = B(x)$.
\item[(2)] \emph{Locality}: the support (in the sense of distributions) of
every $Z^{(n)}\bigl(B_1(x_1)\ox\cdots\bigr)$ lies on the thin diagonal \eqref{eq:diagonal}, 
that is,
$$
\supp Z^{(n)}\bigl(B_1(x_1)\oxyox B_n(x_n)\bigr)\subseteq\Dl_n\ .
$$
\item[(3)] \emph{Field Independence}: 
$$
\fd{Z^{(n)}\bigl(B_1(x_1)\oxyox B_n(x_n)\bigr)}{z}=\sum_{j=1}^n
Z^{(n)}\bigl(B_1(x_1)\oxyox \fd{B_j(x_j)}{z}\oxyox B_n(x_n)\bigr)
$$
and similarly for $\frac{\dl}{\dl\phi^*(z)}$ and $\frac{\dl}{\dl A^\mu(z)}$.
This property is equivalent to the validity of the (causal) Wick expansion
for $Z^{(n)}$.
\item[(4)] \emph{Poincar\'e Covariance}:
$$
\bt_{\La,a} Z^{(n)}\bigl(B_1(x_1)\oxyox B_n(x_n)\bigr) = 
Z^{(n)}\bigl(\bt_{\La,a} B_1(x_1)\oxyox \bt_{\La,a} B_n(x_n)\bigr)\quad
\forall (\La,a)\in\sP_+^\up.
$$
\item[(5)] \emph{$*$-Structure}: 
$$
Z^{(n)}\bigl(B_1(x_1)\oxyox B_n(x_n)\bigr)^*=
Z^{(n)}\bigl(B_1^*(x_1)\oxyox B_n^*(x_n)\bigr).
$$
\item[(6)] \emph{Scaling Degree}: introducing  
$$
z^{(n)}(B_1,\ldots,B_n)(x_1-x_n,\ldots):=\om_0\Bigl(
Z^{(n)}\bigl(B_1(x_1)\oxyox B_n(x_n)\bigr)\Bigr)\in\sD'(\bM^{n-1},\bC)
$$
in analogy to $t_n$ \eqref{eq:tn}, the condition is that
$$
\sd z^{(n)}(B_1,\ldots,B_n)(x_1-x_n,\ldots)\leq\sum_{j=1}^n\dim B_j
$$
for all $B_1,\ldots,B_n\in\sP_\homog$. 
\end{itemize}
\end{defn}

From the property (2) Locality it follows that $\supp
  z^{(n)}(B_1,\ldots,B_n)(x_1-x_n,\ldots)\subseteq\set{(0,\dots,0)}$ and taking 
also into account the property (6) Scaling Degree we conclude that
\begin{align}
&z^{(n)}(B_1,\ldots,B_n)(x_1-x_n,\ldots)=\sum_{|a|=0}^{\om(B_1,\ldots,B_n)}
C_a(B_1,\ldots,B_n)\,\del^a\dl(x_1-x_n,\ldots,x_{n-1}-x_n),\nonumber\\
&\text{with}\quad\om(B_1,\ldots,B_n):=\sum_{j=1}^n\dim B_j-4(n-1)
\label{eq:z}
\end{align}
and some coefficients $C_a(B_1,\ldots,B_n)\in\bC$ depending on 
$B_1,\ldots,B_n$.

\medskip

Denoting the coupling constant by $\ka$ (e.g., $\ka =e$ for scalar QED), let
$\sum_{j=1}^JB_j(g_j)\in\sF_\loc\pw{\ka,\hbar}$
\footnote{$\sF_\loc\pw{\ka,\hbar}$ and $\sP\pw{\ka,\hbar}$ are the vector
spaces of formal power series in the coupling constant $\ka$ and in $\hbar$, 
with coefficients in $\sF_\loc$ and in $\sP$, respectively.}
(with $B_j\in\sP\pw{\ka,\hbar}$ and $g_j\in\sD(\bM)$)
be the interaction,  a typical example is  $S=e\,(j^\mu A_\mu)(g)+e^2\,(A^\mu A_\mu\phi^*\phi)(g^2)$  \eqref{eq:interaction}.
The renormalization of this interaction 
given by the sequence of maps $Z\equiv(Z^{(n)})\in\sR$ is defined by their generating functional
\begin{align}\label{eq:renorm-int}
\sZ\bigl(&(g_j,B_j)_{j=1}^J\bigr):=\sum_{j=1}^JB_j(g_j)\nonumber\\
&+\sum_{n=2}^\infty \frac{1}{n!}\int dx_1\cdots dx_n\,\,\sum_{j_1,\ldots,j_n=1}^J
g_{j_1}(x_1)\cdots g_{j_n}(x_n)\, Z^{(n)}\bigl(B_{j_1}(x_1),\dots,B_{j_n}(x_n)\bigr)
\nonumber\\
=:&\sum_{k=1}^KP_k(f_k)\in\sF_\loc\pw{\ka,\hbar},
\end{align}
by integrating out the $\dl$-distributions appearing in \eqref{eq:z},
where $P_k\in\sP\pw{\ka,\hbar}$ and $f_k\in\sD(\bM)$ are uniquely determined.

The Main Theorem of Renormalization is due to Popineau and Stora \cite{PS16}; 
the more elaborated version given here is essentially taken from \cite{DF04}, 
see also \cite[Chap.~3.6.1-2]{D19} and \cite{BDF09}.
In the formalism at hand, it can be formulated as follows:

\begin{thm}[Main Theorem of Renormalization] \label{th:main-thm-renorm}
\begin{enumerate}
\item[\textup{(a)}]
Given two time-ordered products $T=(T_n)$ and $\wh T=(\wh T_n)$  (with generating functionals
$\nS$ and $\wh\nS$, resp.) both fulfilling 
the axioms (i)--(viii), there exists a unique
renormalization map $Z\in\sR$ fulfilling
\be\label{eq:main-theorem}
\wh\nS \bigl((g_j,B_j)_{j=1}^J\bigr)= \nS\bigl((f_k,P_k)_{k=1}^K\bigr),
\quad\forall B_j\in\sP\pw{\ka,\hbar},\,\,g_j\in\sD(\bM),\,\,J\in\bN,
\ee
where $(f_k,P_k)_{k=1}^K$ is defined in terms of $(g_j,B_j)_{j=1}^J$ and $Z$
according to \eqref{eq:renorm-int}.

\item[\textup{(b)}]
Conversely, given a time-ordered product $T$ fulfilling the axioms (i)--(viii) and an 
arbitrary $Z \in \sR$, the sequence of maps $\wh T\equiv(\wh T_n)_{n=1}^\infty$ 
defined by \eqref{eq:main-theorem} (written concisely in \eqref{eq:hatS=SZ}) satisfies also the axioms
(i)--(viii) for a time-ordered product.
\end{enumerate}
\end{thm}

Interpreting, by abuse of notation, the renormalization of the interaction $\sZ$ \eqref{eq:renorm-int} as the map 
$\sZ:(g_j,B_j)_{j=1}^J\to(f_k,P_k)_{k=1}^K$, the crucial relation \eqref{eq:main-theorem} can be written in a
more catchy form as
\be\label{eq:hatS=SZ}
\wh\nS=\nS\circ\sZ.
\ee

If one selects from the relation \eqref{eq:main-theorem} the terms of 
order $n$ in the $B_j$'s for a $Z\in\sR$ satisfying $Z^{(k)}=0\,\,\forall k\geq 3$
(as it holds for $Z_c$ \eqref{eq:Zk=0}), then one obtains precisely
the equation \eqref{eq:MainThm-n}.

In this paper, we only prove part (b) of this Theorem and only for the 
particular family of elements $Z_c$ of the SP-RG, given in 
\eqref{eq:Z2-general}--\eqref{eq:Zk=0}; this is done in section \ref{sec:hat-T}.

\subsection{Verification that the concretely given $Z_c$ lies in $\sR$}
\label{sec:Z2inR}

$Z_c^{(1)}$ is uniquely determined by the defining property (1) of the SP-RG
$\sR$. In this section, we verify that $Z_c^{(2)}$, concretely given in 
\eqref{eq:Z2-general}--\eqref{eq:zeta}, satisfies the defining properties for 
$Z_c^{(n)}$ given above; this implies then that $Z_c:=(Z_c^{(1)},Z_c^{(2)},0,0,\ldots)$
lies indeed in $\sR$.

Obviously, for any $h\in\sD(\bM^2)$ it holds that
$$
\int dx_1dx_2\,\,h(x_1,x_2)\,Z_c^{(2)}\bigl(B_1(x_1)\ox B_2(x_2)\bigr)
=c\int dx\,\,h(x,x)\,\zeta(B_1,B_2)(x)\word{lies in $\sF_\loc$.}
$$
Linearity, Symmetry \eqref{eq:Z-sym} and Locality (defining property (2)) of $Z_c^{(2)}$ are obvious.
 
To prove the property Field Independence of $Z_c^{(2)}$ (defining property (3)), 
we use the assumption \eqref{eq:assume}, which, e.g.,~implies 
$\fd{}{z}\frac{\del B_j}{\del(\del^\mu\phi)}(x_1)=\frac{\del^2 B_j}{\del(\del^\mu\phi)\,\del\phi}(x_1)\,\dl(x_1-z)$:
\begin{align*}
&\fd{Z_c^{(2)}\bigl(B_1(x_1)\ox B_2(x_2)\bigr)}{z}\\
&\quad =c\,\dl(x_1-x_2,x_1-z)\,
\Bigl(\frac{\del^2 B_1}{\del(\del^\mu\phi^*)\,\del\phi}
\,\frac{\del B_2}{\del(\del_\mu\phi)}+
\frac{\del^2 B_1}{\del(\del^\mu\phi)\,\del\phi}
\,\frac{\del B_2}{\del(\del_\mu\phi^*)}+(B_1\leftrightarrow B_2)\Bigr)(x_1)\\
&\quad =\dl(x_1-z)\,Z_c^{(2)}\Bigl(\frac{\del B_1}{\del\phi}(x_1)\ox B_2(x_2)\Bigr)
+\dl(x_2-z)\,Z_c^{(2)}\Bigl(B_1(x_1)\ox\frac{\del B_2}{\del\phi}(x_2)\Bigr)\\
&\quad =Z_c^{(2)}\Bigl(\fd{B_1(x_1)}{z}\ox B_2(x_2)\Bigr)
+Z_c^{(2)}\Bigl(B_1(x_1)\ox\fd{B_2(x_2)}{z}\Bigr)
\end{align*}
and similarly for $\frac{\dl}{\dl\phi^*(z)}$ and $\frac{\dl}{\dl A^\mu(z)}$. In the 
last step we have used again \eqref{eq:assume} to conclude that, e.g.,~the term 
$\frac{\del B_1}{\del(\del^\nu\phi)}(x_1)\,\del^\nu\dl(x_1-z)$ of $\fd{B_1(x_1)}{z}$ does not 
contribute to $Z_c^{(2)}\Bigl(\fd{B_1(x_1)}{z}\ox B_2(x_2)\Bigr)$.

Poincar\'e Covariance (defining property (4)):
Translation covariance of  $Z_c^{(2)}$ is obvious and Lorentz covariance
follows from the fact that $\zeta(B_1,B_2)(x_1)$ is a Lorentz tensor of 
the same type as $B_1(x_1)\,B_2(x_2)$.

The property $*$-Structure of $Z_c^{(2)}$ (defining property (5)) follows from 
$\zeta(B_1,B_2)^*=\zeta(B_1^*,B_2^*)$, which relies on
$\Bigl(\frac{\del B}{\del(\del^\mu\phi)}\Bigr)^*=
\frac{\del B^*}{\del(\del^\mu\phi^*)}$.

To verify the property Scaling Degree (defining property (6)) note first that $z_c^{(2)}(B_1,B_2)$
is non-vanishing only for $(B_1,B_2)=(\del^\mu\phi,\del^\nu\phi^*)$
or $(B_1,B_2)=(\del^\nu\phi^*,\del^\mu\phi)$. In both cases it holds that
$\zeta(B_1,B_2)=g^{\mu\nu}$, so we obtain
$$
\sd z_c^{(2)}(B_1,B_2)(y)=\sd(g^{\mu\nu}\,\dl(y))=4=
\dim\del^\mu\phi+\dim\del^\nu\phi^*.
$$

\subsection{Proof that $\wh T_c$ constructed from $T$ and $Z_c$ by
\eqref{eq:MainThm-n} is a time-ordered product}\label{sec:hat-T}
 
In this section we prove that $\wh T_c$, defined in \eqref{eq:MainThm-n} in 
terms of $T$ and the concretly given $Z_c$, satisfies the basic axioms (i)-(iv) and the
renormalization conditions (v)-(viii) given in Appendix \ref{sec:axioms-T}. 
This statement is part (b) of the Main Theorem
for the particular $Z_c$ given in \eqref{eq:Z2-general}--\eqref{eq:Zk=0}.
Since, in contrast to \cite[Chapt.~3.6.1-2]{D19} and \cite{DF04,BDF09}, 
we are forced to work in a formalism not fulfilling the AWI, 
we cannot refer to the general proof of the Main Theorem given in these 
references.

\paragraph{Basic axioms.} 
The Initial Condition (ii), $\wh T_{c,1}(B(x))=B(x)$, is obvious.
Linearity (i) in $B_1\oxyox B_n$ and Symmetry (iii) follow from the corresponding 
properties of $T$ and $Z_c$, as we see by looking at \eqref{eq:MainThm-n}.

To verify Causality (iv) let 
$\set{x_1,\ldots,x_k}\cap(\set{x_{k+1},\ldots,x_n}+\ovl V_-)=\emptyset$. By
Locality of $Z_c^{(2)}$ (defining property (2)) it holds that
$$
Z_c^{(2)}\bigl(B_j(x_j)\ox B_l(x_l)\bigr)=0
\word{if $1\leq j\leq k$ and $k+1\leq l\leq n$.}
$$
Using this and in a second step Causality (iv) of $T$ we indeed obtain
causal factorization of $\wh T_{c,n}$, in detail:
\begin{align*}
i^n\,&\wh T_{c,n}\bigl(\ox_{j=1}^n B_j(x_j)\bigr)=
\sum_{\substack{P\in\mathrm{Part}_2(\{1,\dots,k\})\\ k/2\leq|P|\leq k}}\,\,
\sum_{\substack{Q\in\mathrm{Part}_2(\{k+1,\dots,n\})\\(n-k)/2\leq|Q|\leq n-k}} i^{|P|+|Q|}\\
&\quad \cdot T_{|P|+|Q|}\Bigl(
\bigox_{I\in P} Z_c^{(|I|)}\bigl(\ox_{j\in I}B_j(x_j)\bigr)\ox
\bigox_{R\in Q} Z_c^{(|R|)}\bigl(\ox_{r\in R}B_r(x_r)\bigr)\Bigr)\\
&=\sum_{\substack{P\in\mathrm{Part}_2(\{1,\dots,k\})\\ k/2\leq|P|\leq k}}
i^{|P|}\,T_{|P|}\Bigl(\bigox_{I\in P} Z_c^{(|I|)}\bigl(\ox_{j\in I}B_j(x_j)\bigr)\Bigr)\\
&\qquad\qquad\qquad\qquad
\star\sum_{\substack{Q\in\mathrm{Part}_2(\{k+1,\dots,n\})\\(n-k)/2\leq|Q|\leq n-k}}
i^{|Q|}\,T_{|Q|}\Bigl(\bigox_{R\in Q} Z_c^{(|R|)}\bigl(\ox_{r\in R}B_r(x_r)\bigr)\Bigr)\\
&=i^n\,\wh T_{c,k}\bigl(\ox_{j=1}^k B_j(x_j)\bigr)\star 
\wh T_{c,n-k}\bigl(\ox_{r=k+1}^n B_r(x_r)\bigr).
\end{align*}

\paragraph{Renormalization conditions.} The validity of axioms Field Independence (v)
and Poincar\'e Covariance (vii) for $\wh T_c$ follows straightforwardly from the 
corresponding properties of $T$ and $Z_c$. 

To verify the axiom (vi) $*$-Structure for $\wh T_c$, we first conclude from the
property $*$-Structure of $Z_c$ (defining property (5)) and \eqref{eq:renorm-int} that, if 
$\sZ_c\bigl((g_j,B_j)_j\bigr)=(f_k,P_k)_k$ (by abuse of notation), then
$\sZ_c\bigl((\ovl{g_j},B_j^*)_j\bigr)=(\ovl{f_k},P_k^*)_k$. With this and by using \eqref{eq:hatS=SZ} 
(or \eqref{eq:main-theorem}, resp.) and the axiom $*$-Structure  for $T$, we obtain
$$
\wh\nS_c\bigl((\ovl{g_j},B_j^*)_j\bigr)^{\star -1}=\nS\bigl(\sZ_c((\ovl{g_j},B_j^*)_j)\bigr)^{\star -1}
=\nS\bigl((\ovl{f_k},P_k^*)_k\bigr)^{\star -1}=\nS\bigl((f_k,P_k)_k\bigr)=\wh\nS_c\bigl((g_j,B_j)_j\bigr).
$$
To prove that $\wh T_c$ satisfies the axiom (viii) Scaling Degree, first note that, up 
to permutations of $(B_1(x_1),\ldots,B_n(x_n))$ and the prefactor $i^{|P|}$, 
every summand of
$$
\wh t_{c,n}(B_1,\ldots,B_n)(x_1-x_n,\ldots)=
\sum_{\substack{P\in\mathrm{Part}_2(\{1,\dots,n\})\\ n/2\leq|P|\leq n}}i^{|P|}\,
\om_0\left(T_{|P|}\Bigl(\bigox_{I\in P} Z_c^{(|I|)}\bigl(\ox_{j\in I}B_j(x_j)\bigr)
\Bigr)\right)
$$
is equal to
\begin{align}
&\om_0\Bigl(T_{n-r}\Bigl(\bigox_{j=1}^r Z_c^{(2)}\bigl(B_j(x_j)\ox B_{r+j}(x_{r+j})
\bigr)\ox\bigox_{s=2r+1}^nB_s(x_s)\Bigr)\Bigr)\nonumber\\
&=c^r\,t_{n-r}\bigl(\zeta(B_1,B_{r+1}),\ldots,
\zeta(B_r,B_{2r}),B_{2r+1},\ldots,B_n\bigr)
(x_1-x_n,\ldots,x_r-x_n,x_{2r+1}-x_n,\ldots)\nonumber\\
&\quad\cdot\prod_{j=1}^r\dl(x_j-x_{r+j})\label{eq:VEV}
\end{align}
for some $0\leq r\leq n/2$. Next note that for $B_1,B_2\in\sP_\homog$ it
holds that $\zeta(B_1,B_2)\in\sP_\homog$ and that
$$
\dim\zeta(B_1,B_2)\leq\dim B_1+\dim B_2-4.
$$
Using additionally the axiom (viii) Scaling Degree for $T$ and the formulas 
$\sd \dl(x_j-x_{r+j})=4$ and $\sd(f_1\ox f_2)=\sd(f_1)+\sd(f_2)$, we 
see that the scaling degree of the expression on the r.h.s.\ of 
\eqref{eq:VEV} is bounded by
$$
\sd(\ldots)\leq\sum_{j=1}^r\dim\zeta(B_j,B_{j+r})+\sum_{s=2r+1}^n\dim B_s
+4r\leq\sum_{j=1}^n\dim B_j.
$$

\paragraph{Acknowledgements.} We thank the referee for reading the manuscript
extremely thoroughly and pointing out a lot of improvements. M.D.~profited from
enlightening discussions with Klaus Fredenhagen, Romeo Brunetti and Kasia Rejzner.


\begin{thebibliography}{10}\itemsep0mm

\bibitem{BS59}
N.N.~Bogoliubov and D.V.~Shirkov,
\emph{Introduction to the Theory of Quantized Fields},
Interscience Publishers, 1959.

\bibitem{BD08} 
F.~Brennecke and M.~D\"utsch,
``Removal of violations of the Master Ward Identity
in perturbative QFT'',
Rev. Math. Phys. \textbf{20} (2008), 119--172.

\bibitem{BDF09} 
R.~Brunetti, M.~D\"utsch and K.~Fredenhagen,
``Perturbative algebraic quantum field theory and the renormalization 
groups'',
Adv. Theor. Math. Phys. \textbf{13} (2009), 1541--1599.

\bibitem{BDFR20} 
R.~Brunetti, M.~D\"utsch, K.~Fredenhagen and K.~Rejzner,
work in progress.

\bibitem{BF19}
D.~Buchholz and K.~Fredenhagen,
``A $C^*$-algebraic Approach to Interacting Quantum Field Theories'', 
Commun. Math. Phys. \textbf{377} (2020), 947--969.

\bibitem{D19}
M.~D\"utsch,
``From Classical Field Theory to Perturbative Quantum Field Theory'',
Progress in Mathematical Physics \textbf{74}, Birkh\"auser, 2019.

\bibitem{DB02} 
M.~D\"utsch and F.-M.~Boas,
``The Master Ward Identity'',
Rev. Math. Phys. \textbf{14} (2002), 977--1049.

\bibitem{DF99} 
M.~D\"utsch and K.~Fredenhagen,
``A local (perturbative) construction of observables in gauge
theories: the example of QED'',
Commun. Math. Phys. \textbf{203} (1999), 71--105.

\bibitem{DF03} 
M.~D\"utsch and K.~Fredenhagen,
``The Master Ward Identity and generalized Schwinger--Dyson equation
in classical field theory'',
Commun. Math. Phys. \textbf{243} (2003), 275--314.

\bibitem{DF04} 
M.~D\"utsch and K.~Fredenhagen,
``Causal perturbation theory in terms of retarded products,
and a proof of the Action Ward Identity'',
Rev. Math. Phys. \textbf{16} (2004), 1291--1348.

\bibitem{DKS93}
M.~D\"utsch, F.~Krahe and G.~Scharf,
``Scalar QED Revisited'',
Nuovo Cimento \textbf{A 106} (1993), 277--307. 

\bibitem{EG73}
H.~Epstein and V.~Glaser,
``The role of locality in perturbation theory'',
Ann. Inst. Henri Poincar\'e \textbf{19A} (1973), 211--295.

\bibitem{P20}
L.~Peters,
``The Master Ward Identity for the complex scalar field: From classical
to quantum symmetries'', Bachelor's Thesis, G\"ottingen
University, 2020, 
arXiv:2103.05433

\bibitem{PS16}
G.~Popineau and R.~Stora,
``A pedagogical remark on the main theorem of perturbative
renormalization theory'', 
Nucl. Phys. B \textbf{912} (2016), 70--78, 
preprint: LAPP--TH, Lyon (1982).

\bibitem{R20} 
K.~Rejzner,
``BV quantization in perturbative algebraic QFT: Fundamental 
concepts and perspectives'',
arXiv:2004.14272 (2020)

\bibitem{Tipp19} F.~Tippner, ``Scalar QED with
String-Localised Potentials'', Bachelor's Thesis, G\"ottingen
University, 2019.
\end{thebibliography}
\end{document}